\documentclass[11pt]{article}

\usepackage{fullpage}
\usepackage{latexsym}
\usepackage[english]{babel}

\usepackage{amsfonts,amsmath,amssymb,amsthm,mathtools}
\usepackage{xspace,graphicx,relsize,bm,xcolor,mathrsfs}
\usepackage{dsfont}
\usepackage{booktabs}
\usepackage{array}
\usepackage{siunitx}
\usepackage{microtype}

\usepackage[numbers,sort&compress]{natbib}
\usepackage{hyperref}
\usepackage{orcidlink}

\hypersetup{
    colorlinks,
    linkcolor={blue!100!black},
    citecolor={blue!100!black},
    urlcolor={blue!100!black}
}

\newtheorem{theorem}{Theorem}

\newtheorem{proposition}[theorem]{Proposition}
\newtheorem{corollary}[theorem]{Corollary}

\theoremstyle{definition}

\theoremstyle{remark}

\newenvironment{frontmatter}
  {}
  {\par\bigskip}

\renewenvironment{abstract}
  {%
    \par\bigskip
    \begin{center}
      \bfseries Abstract
    \end{center}
    \noindent
  }
  {%
    \par\bigskip
  }

\newenvironment{keyword}
  {%
    \par\smallskip
    \noindent{\bfseries Keywords: }
  }
  {%
    \par\bigskip
  }

\newcommand{\sep}{\unskip; }

\title{\bfseries
SPID-Chain: Verifiable Polar-Coded State Validation\\
for Cross-Chain DAG Settlement
}

\author{%
Amirhossein Taherpour\texorpdfstring{\,\orcidlink{0000-0003-4647-102X}}{}
\quad
Xiaodong Wang\texorpdfstring{\,\orcidlink{0000-0002-2945-9240}}{}\\[0.5em]
Department of Electrical Engineering, Columbia University\\
500 West 120th Street, New York, NY 10027, USA\\[0.4em]
\textit{Email addresses:}
\href{mailto:at3532@columbia.edu}{at3532@columbia.edu};
\href{mailto:xw2008@columbia.edu}{xw2008@columbia.edu}
}

\date{}

\begin{document}
\pagestyle{plain}
\pagenumbering{arabic}
\selectlanguage{english}

\begin{frontmatter}

\maketitle

\begin{abstract}
Cross-chain settlement must preserve safety across heterogeneous ledgers while tolerating delayed computation, Byzantine participants, and adversarial transaction issuance. This paper presents SPID-Chain, an adapter-compatible settlement architecture for escrow-backed fungible transfers across programmable blockchains. SPID-Chain maintains settlement state through persistent Polar-coded fragments, validates candidate state transitions using hidden linear verification checks, and records certified transfers in a weighted directed acyclic graph (DAG). The design separates native-chain finality from cross-chain settlement: source-chain finality establishes an immutable reservation, whereas weighted DAG confirmation determines when the corresponding destination credit becomes executable. We derive an exact recovery-time distribution for heterogeneous coded workers, a verification-soundness bound for Byzantine responses, and an exact weighted-quorum condition for conflicting-block safety. These components are coupled in a cross-layer stability theorem showing how the coded-validation completion probability determines the effective honest issuance rate and, consequently, the stable adversarial-load region of the settlement DAG. We further establish an end-to-end settlement guarantee covering balance non-negativity, asset conservation, conflict exclusion, replay protection, coded-state consistency, and finite expected lock-to-release latency under the stated liveness conditions. Prototype-assisted simulations indicate that coded validation reduces sensitivity to stragglers, improves validation and confirmation throughput under heterogeneous delays, and produces the predicted transition between stable and unstable DAG operation. The resulting framework provides a verifiable and analytically grounded settlement layer without modifying the native consensus protocol of participating chains.
\end{abstract}

\begin{keyword}
Cross-chain settlement \sep
coded computing \sep
Polar codes \sep
directed acyclic graph \sep
Byzantine verification
\end{keyword}

\end{frontmatter}

\section{Introduction}
\label{sec:introduction}

Blockchain ecosystems increasingly comprise multiple independent execution
and settlement domains. Participating chains may differ in their consensus
protocols, validator sets, state representations, finality rules, transaction
semantics, and performance characteristics. Applications that transfer assets
or coordinate state across these domains must therefore establish more than
the existence of a message on a remote chain. They must determine that the
source event is final, the corresponding funds have been reserved, the
proposed state transition is admissible and replay-free, conflicting
transitions cannot both succeed, and the destination effect is executed
exactly once. Providing these guarantees without requiring participating
chains to replace their native consensus protocols remains a central problem
in blockchain interoperability
\cite{Belchior2021InteropSurvey,Ren2023InteropTKDE,
Ou2022CrossChainOverview,Li2025CrossChainSurvey,
Zamyatin2021SoKCommunication}.

A broad range of interoperability mechanisms has been developed to address
this problem. Atomic-swap protocols coordinate asset exchange through
cryptographic and timing conditions, while sidechains, relays, and
light-client constructions authenticate the evolution of remote ledgers.
Collateral-backed protocols, notary systems, gateways, and middleware
architectures coordinate transfers through external participants and
recovery procedures. Proof-carrying bridges and zero-knowledge constructions
instead verify remote consensus or state transitions through succinct
cryptographic evidence. Other approaches combine the trust of multiple
blockchains, support multi-party asset movement, or standardize transaction
consistency across heterogeneous ledgers
\cite{Herlihy2018AtomicSwaps,Zamyatin2019XCLAIM,
Gazi2019PoSSidechains,Kiayias2020PoWSidechains,
Garoffolo2020Zendoo,Xie2022zkBridge,Belchior2022Hermes,
Sheng2023TrustBoost,Augusto2024MultiParty,Kumar2025TwinToken,
IEEE3221012025}. Complementary security and empirical studies examine bridge
attack surfaces, inconsistent cross-chain outcomes, operational costs,
economic effects, and the distinction between nominal interoperability and
the cross-chain activity realized in deployed systems
\cite{Augusto2024SoKSecurityPrivacy,Li2025BridgeSecurity,
Yan2025CrossChainTransactions,Cao2026PriceInterop}.

Directed acyclic graph (DAG)-based ledgers constitute another relevant
research cluster. By allowing multiple blocks or transactions to be issued
concurrently and related through references, a DAG can avoid immediate
serialization into a single global chain. Existing work studies DAG
structures, weighted or reference-based confirmation, parent-selection
policies, tip-pool stability, transaction attachment, and end-to-end
performance
\cite{Wang2023DAGSoK,Muller2022Tangle20,
Muller2023TipPoolStability,Guo2025DAGAttachment,
Shi2025DAGPerformance,Taherpour2025CodedBlockchainIoT,Taherpour2024HybridChain,Taherpour2026ZKHybridFL,Taherpour2026GossipVirtualVoting}. Such mechanisms are attractive for an external
settlement layer because blocks issued by different chains can concurrently
approve earlier cross-chain transfers and contribute distinct-chain support
toward their confirmation.

A parallel body of work addresses the computational cost of distributed
validation. Coded distributed computing introduces structured redundancy so
that a coordinator can reconstruct a result from a decodable subset of
worker responses rather than waiting for every assigned worker. This
principle has been developed through maximum-distance-separable codes, binary
linear codes, Polar codes, computational polarization, and other
straggler-resilient constructions
\cite{Ng2021CDCsurvey,Lee2018SpeedingUp,Reed1960PolynomialCodes,
Arikan2009ChannelPolarization,Bartan2019PolarServerless,
Soleymani2021BinaryLinearCodes,Pilanci2022ComputationalPolarization,
Fathollahi2022PolarScaling}. Secure and verifiable coded-computing research
further considers erroneous or adversarial worker responses, privacy, and
the separation of computation recovery from result verification
\cite{Yu2019LagrangeCoded,Tang2022AVCC,Hong2024GVCC,
Byrne2023AdversaryTolerant,Karpuk2024ModularPolynomial,
Freivalds1979FastProbabilistic,Backes2013VerifiableDelegation}.
Byzantine quorum systems and partially synchronous consensus models,
meanwhile, provide the foundations for excluding conflicting decisions and
separating unconditional safety from eventual progress
\cite{Malkhi1998ByzantineQuorum,Dwork1988PartialSynchrony}.

Across these research clusters, the decisive unresolved issue is the
cross-layer dependence between distributed validation and cross-chain
confirmation. Interoperability protocols principally specify how remote
events are authenticated, coordinated, or applied. Coded-computing systems
principally analyze the recovery of individual distributed computations. DAG
analyses generally begin from an exogenous block-issuance or approval
process, while quorum and verification mechanisms establish local integrity
or decision conditions. These views do not characterize a settlement layer in
which a persistently coded shared state is validated by heterogeneous and
potentially Byzantine workers, the probability of completing that validation
determines the effective issuance rate of honest settlement blocks, and that
issuance rate in turn determines the stability and confirmation behavior of a
weighted cross-chain DAG. Consequently, the relationship among code
configuration, worker delays, verification overhead, DAG parent selection,
adversarial issuance, and end-to-end settlement safety remains
uncharacterized.

This paper presents SPID-Chain, an adapter-compatible architecture for
escrow-backed fungible settlement across programmable blockchains.
SPID-Chain separates native-chain finality from cross-chain confirmation.
Native consensus finalizes a source reservation, thereby establishing that
the outgoing funds are immutably locked, but the associated destination
credit is not released until the resulting cross-chain block has been
validated and confirmed by the external settlement layer. This separation
allows participating chains to retain their existing consensus protocols
while exposing a common adapter interface for finalized reservations,
sequence numbers, replay information, confirmation certificates, and
destination application.

For each outgoing instance, an external committee jointly validates the local
transfer batch and a reproducibly selected set of candidate DAG parents
against a confirmed settlement checkpoint. Physical workers maintain
persistent Polar-coded fragments of the checkpoint state and return coded
results for the complete validation workload. The committee accepts a
response only after signature, commitment, and hidden linear verification
checks succeed; missing and rejected responses are treated as erasures, and
the requested state values are recovered once the accepted response pattern
has full rank. A successfully validated transfer is then recorded in a
weighted settlement DAG. Descendant blocks issued by distinct chains
contribute their protocol-level weights to the blocks in their past cones,
and a transfer becomes executable only after its accumulated support reaches
the confirmation threshold. Confirmed blocks are applied in a deterministic
checkpoint order, after which both the raw settlement state and its
persistent coded representation are updated incrementally.

This construction exposes a cross-layer tradeoff that is absent when the
validation and confirmation layers are considered separately. Increasing the
DAG parent budget allows an honest block to approve more existing tips and
can enlarge the stable adversarial-load region. However, every candidate
parent is also an additional state-validation item, so a larger parent budget
increases worker computation, communication, and verification costs and can
reduce the probability that an honest proposal completes before its
deadline. Similarly, increasing coding redundancy improves tolerance to
missing responses but increases storage and processing overhead. SPID-Chain
therefore analyzes the coding rate, validation deadline, parent budget, and
adversarial issuance rate as coupled protocol parameters rather than as
independent design choices.

The scope of the present model is deliberately narrower than universal
cross-chain execution. SPID-Chain considers deterministic linear settlement
updates induced by escrow-backed fungible transfers, assumes that each issuer
has at most one outstanding outgoing sequence, and requires every honest
supporting chain to reproduce the deterministic admissibility checks of the
block it supports. These restrictions make the interaction among source
reservations, persistent coded state, weighted confirmation, and
exactly-once destination execution explicit and amenable to rigorous
analysis. The protocol does not claim to support arbitrary cross-chain
smart-contract execution or to correct failures below the participating
chains' adapter interfaces.

The principal contributions are as follows.
\begin{enumerate}
    \item We specify an adapter-compatible cross-chain settlement protocol
    that separates native source-chain finality from cross-chain
    confirmation. The protocol integrates finalized source reservations,
    deterministic replay protection, committee-certified validation,
    reproducible DAG parent selection, weighted distinct-chain confirmation,
    deterministic checkpoint application, and exactly-once destination
    execution without replacing the native consensus protocol of a
    participating chain.

    \item We develop a persistent Polar-coded validation layer for
    heterogeneous workers. The construction supports reliability-aware
    information-set and worker placement, incremental coded-state updates,
    joint validation of outgoing transfers and candidate DAG parents,
    erasure treatment of missing or rejected responses, rank-based recovery,
    and hidden linear checks for detecting Byzantine results.

    \item We derive an exact recovery-time distribution for heterogeneous
    coded workers, a verification-soundness bound for Byzantine responses,
    and an exact weighted-quorum condition for conflicting-block safety,
    together with parameter-selection corollaries for verification checks and
    uniform-weight deployments. Our principal cross-layer result shows how
    the coded-validation completion probabilities determine the effective
    honest block-issuance rate and, through that rate, the stability
    boundary of the settlement DAG. We further establish an end-to-end
    settlement guarantee covering balance non-negativity, asset conservation,
    conflict exclusion, replay protection, exactly-once destination
    application, coded-state consistency, and finite expected
    lock-to-release latency under the stated liveness conditions.

    \item We evaluate SPID-Chain through a prototype-assisted discrete-event
    simulation comparing Polar-coded validation with uncoded execution,
    replication, and maximum-distance-separable coding. The experiments
    examine heterogeneous worker delays, deadline completion, validation and
    confirmation throughput, worker and chain population, event-certified
    execution, committee-to-worker ratio, long-run DAG stability, and the
    joint effect of the parent budget and adversarial issuance.
\end{enumerate}

The remainder of the paper is organized as follows.
Section~\ref{sec:related-work} reviews cross-chain interoperability,
DAG-based settlement, coded distributed computing, and Byzantine
verification. Section~\ref{sec:system-model} defines the system, fault, and
settlement models. Section~\ref{sec:spid-protocol} presents the SPID-Chain
architecture and protocol. Section~\ref{sec:polar-validation} develops the
persistent Polar-coded validation mechanism.
Section~\ref{sec:security} establishes the correctness and cross-layer
performance results. Section~\ref{sec:evaluation} describes the simulation
methodology and evaluation. Section~\ref{sec:discussion} discusses the
assumptions, deployment implications, and limitations, and
Section~\ref{sec:conclusion} concludes the paper.

\section{Related Work}
\label{sec:related-work}

SPID-Chain intersects four research areas: blockchain interoperability,
directed acyclic graph (DAG)-based ledgers, coded distributed computing, and
Byzantine verification. This section reviews the closest work in each area
and identifies the cross-layer problem that remains unresolved.

\subsection{Cross-Chain Interoperability and Settlement}
\label{subsec:related-cross-chain}

Blockchain interoperability mechanisms have been classified according to
their communication model, verification method, trust assumption, and
supported state transition. Representative classes include atomic swaps,
notary and gateway systems, sidechains, relays, light clients, and
proof-carrying bridges
\cite{Belchior2021InteropSurvey,Ren2023InteropTKDE,
Ou2022CrossChainOverview,Li2025CrossChainSurvey}. Comparative studies further
show that interoperability cannot be assessed solely by whether two ledgers
can exchange messages; transaction atomicity, remote-state authenticity,
fault isolation, replay resistance, and recovery from interrupted executions
are equally important
\cite{Koens2019AssessingInterop,Hardjono2020BlockchainAS}. The publication of
IEEE Std~3221.01-2025 reflects the continuing standardization of cross-chain
transaction-consistency mechanisms
\cite{IEEE3221012025}.

Atomic cross-chain swaps use hash and time constraints to exchange assets
without a common ledger \cite{Herlihy2018AtomicSwaps}. Their objective differs
from SPID-Chain because they coordinate an exchange among designated parties
rather than maintain a shared settlement state for continuing transfers.
XCLAIM supports cryptocurrency-backed cross-chain assets through
collateralized intermediaries and explicit financial penalties
\cite{Zamyatin2019XCLAIM}. Proof-of-stake and proof-of-work sidechain
constructions instead authenticate remote-chain evolution through
cross-chain proofs and sidechain-specific security assumptions
\cite{Gazi2019PoSSidechains,Kiayias2020PoWSidechains}.

Proof-oriented systems reduce reliance on external notaries. Zendoo uses
zero-knowledge proofs to decouple sidechain execution from verification on a
main chain \cite{Garoffolo2020Zendoo}, whereas zkBridge verifies remote
consensus and state transitions through succinct proofs
\cite{Xie2022zkBridge}. These systems provide stronger cryptographic
verification of remote ledgers than SPID-Chain's adapter model, but proof
generation and light-client verification solve a different problem from
straggler-resilient evaluation of a shared settlement state.

Gateway and middleware architectures are closer to the deployment model
considered here. HERMES extends the Open Digital Asset Protocol with
fault-tolerant logging and recovery for gateway-mediated transfers
\cite{Belchior2022Hermes}. More recent work studies multi-party asset
transfers and gateway-independent token synchronization across heterogeneous
chains
\cite{Augusto2024MultiParty,Kumar2025TwinToken}. These approaches focus on
transaction coordination, gateway recovery, or token representation.
SPID-Chain instead addresses the computational bottleneck created when
multiple candidate transfers must be evaluated consistently against a common
checkpoint before settlement.

TrustBoost studies how several blockchains can combine their security without
changing their underlying consensus protocols
\cite{Sheng2023TrustBoost}. This ``consensus on top of consensus'' direction
is conceptually related to SPID-Chain's use of distinct-chain support.
However, TrustBoost constructs a combined trust layer, whereas SPID-Chain
uses a weighted settlement DAG for escrow-backed transfers and explicitly
couples confirmation capacity to the probability of completing coded
validation.

Bridge security remains a material concern even when the participating
ledgers are individually secure. Recent work catalogues failures arising from
smart-contract defects, compromised validators, inconsistent cross-chain
state, liquidity assumptions, and off-chain coordination
\cite{Li2025BridgeSecurity}. Empirical studies have also measured
cross-chain transaction costs, inconsistent outcomes, attack-related
activity, and the difference between deployed bridge connectivity and
realized cross-chain use
\cite{Yan2025CrossChainTransactions,Cao2026PriceInterop}. These findings
motivate SPID-Chain's explicit separation of source-chain finality,
cross-chain admissibility, confirmation, replay protection, and destination
application. SPID-Chain's guarantees apply to the adapter and adversarial
models stated in Section~\ref{sec:system-model}.

\subsection{DAG-Based Ledgers and Interchain Confirmation}
\label{subsec:related-dag}

DAG-based distributed ledgers allow multiple blocks or transactions to be
issued concurrently and related through references rather than immediately
serialized into one global chain. Existing systems differ in their vertex
semantics, parent-selection policies, ordering rules, conflict resolution,
and confirmation criteria
\cite{Wang2023DAGSoK}. Tangle~2.0, for example, combines a block DAG with
leaderless consensus and weight-based branch selection
\cite{Muller2022Tangle20}.

The size and consistency of the public tip set are central to the performance
of a DAG ledger. Analytical work on Tangle-like systems studies the effect of
issuance rates, communication delays, and local views on tip-pool stability
\cite{Muller2023TipPoolStability}. Recent transaction-attachment algorithms
seek to balance tip selection, scalability, and resistance to lazy-tip or
parasite-chain behavior
\cite{Guo2025DAGAttachment}. Complementary performance studies examine the
complete transaction lifecycle and show that throughput and latency alone do
not capture the influence of DAG topology and processing behavior
\cite{Shi2025DAGPerformance}.

SPID-Chain uses a DAG as an external settlement and confirmation structure,
not as a replacement for native-chain consensus. Its parent-selection model
also differs from protocols in which a transaction approves tips solely to
maintain the native ledger. In SPID-Chain, a candidate parent contributes an
additional state-validation item. Increasing the parent budget $K$ can
therefore improve tip-removal capacity while simultaneously reducing the
probability that validation completes before its deadline. The stability
condition in Theorem~\ref{thm:cross-layer-stability} captures this coupling
through the effective honest issuance rate. This dependence of DAG stability
on coded-validation completion is not represented in prior tip-selection or
DAG-performance models.

\subsection{Coded Distributed Computing}
\label{subsec:related-coded-computing}

Coded distributed computing introduces structured redundancy into worker
assignments so that a coordinator can recover a result without waiting for
every worker. Early constructions established that coding can reduce
straggler-induced latency in distributed linear algebra and machine learning
\cite{Lee2018SpeedingUp}. Subsequent work has characterized computation,
communication, recovery-threshold, privacy, and robustness trade-offs across
a broad range of distributed workloads
\cite{Ng2021CDCsurvey}.

Maximum-distance-separable constructions based on Reed--Solomon codes recover
from any sufficiently large set of returned symbols
\cite{Reed1960PolynomialCodes}. Their response-set condition depends only on
the number of available symbols, but dense finite-field encoding and decoding
can be expensive for persistent state. Polar codes originate from channel
polarization \cite{Arikan2009ChannelPolarization}. Their recursive transforms
have subsequently been adapted to straggler-resilient serverless and
distributed computation
\cite{Bartan2019PolarServerless,Soleymani2021BinaryLinearCodes,
Pilanci2022ComputationalPolarization,Fathollahi2022PolarScaling}. Unlike an
MDS code, a finite-length Polar construction is generally recoverable from
some response sets of a given cardinality but not from all such sets. This is
why the recovery analysis in
Proposition~\ref{prop:recovery-time-law} is expressed through the complete
decodable-set family rather than through a scalar recovery threshold.

Recent coded-computing research has expanded beyond exact straggler recovery.
Learning-theoretic coding supports approximate recovery for more general
computations \cite{Moradi2024LearningTheoretic}, while secure polynomial-code
constructions consider privacy, adversarial results, and robustness
\cite{Byrne2023AdversaryTolerant,Karpuk2024ModularPolynomial}. These
constructions primarily target isolated distributed-computation jobs.
SPID-Chain instead maintains coded fragments persistently across settlement
checkpoints, updates them incrementally after confirmation, and applies the
same encoded state to a joint workload containing the local outgoing batch
and its candidate DAG parents.

\subsection{Verification and Byzantine Quorums}
\label{subsec:related-verification}

Randomized linear verification provides an inexpensive mechanism for testing
the correctness of linear-algebraic computations. Freivalds' method detects
an incorrect matrix product through a randomly selected linear projection
\cite{Freivalds1979FastProbabilistic}. Cryptographic verifiable-computation
systems provide stronger delegation guarantees for general outsourced
computations, but typically introduce additional proof-generation and
verification machinery
\cite{Backes2013VerifiableDelegation}. SPID-Chain adopts a narrower mechanism:
workers commit to their coded responses before independently hidden linear
checks are applied. This produces a field-dependent soundness bound suited to
the linear settlement workload.

At the confirmation layer, Byzantine quorum systems characterize how
intersecting decision sets prevent conflicting outcomes
\cite{Malkhi1998ByzantineQuorum}. Partial synchrony separates unconditional
safety from liveness after network stabilization
\cite{Dwork1988PartialSynchrony}. SPID-Chain specializes these foundations to
heterogeneous chain weights. Its exact condition is expressed through the
minimum weight of the intersection of two confirmation quorums, rather than
through a uniform replica count.

\subsection{Positioning of SPID-Chain}
\label{subsec:related-positioning}

SPID-Chain targets the layer coupling that remains implicit when
interoperability, coded validation, DAG confirmation, and quorum safety are
studied separately. Prior interoperability systems do not model coded worker
completion as a determinant of settlement-DAG capacity; prior
coded-computing systems do not maintain a confirmed cross-chain state whose
updates depend on weighted DAG confirmation; and prior DAG stability
analyses do not incorporate a code-dependent probability that an honest
proposal becomes a valid settlement block. SPID-Chain closes this gap by
connecting the heterogeneous response-set law, verification soundness,
effective honest issuance, DAG stability, weighted confirmation, and
settlement invariants within one protocol model.

\section{System Model and Problem Formulation}
\label{sec:system-model}

\subsection{Cross-Chain State and Execution Model}
\label{subsec:cross-chain-model}

Consider a set of $N$ participating blockchains indexed by
$\mathcal{N}=\{1,\ldots,N\}$. Each chain retains its native consensus
protocol and interacts with SPID through a gateway adapter implemented as a
smart contract, runtime module, or native protocol component. The adapter
exports natively finalized outgoing lock events, verifies cross-chain
certificates, and applies confirmed incoming transfers. Native consensus
therefore finalizes the source-chain lock event; it does not by itself
finalize the corresponding cross-chain settlement.

A participating chain is \emph{adapter-compatible} if its gateway satisfies
the following conditions. Each outgoing event is assigned an issuer-specific
sequence number and is accompanied by a verifiable native-finality
certificate. An honest gateway never finalizes two distinct events with the
same sequence number, and it applies each confirmed incoming transfer at most
once. Participating chains may use different native consensus mechanisms,
provided that they expose this common interface.

Time is divided into checkpoint intervals indexed by $t\geq 0$. Interval $t$
begins from a confirmed cross-chain checkpoint with state commitment
$h^{\mathrm{cp}}(t)$. An honest chain may issue a block in interval $t$ only
after adopting that checkpoint. Hence, all active honest committees in the
same interval validate against the same confirmed settlement state, although
their views of unconfirmed DAG tips may differ because of communication
delay.

For chain $j$, let $\mathcal{C}_j(t)$ and $\mathcal{W}_j(t)$ denote its
committee and physical-worker sets during interval $t$, respectively, and let
$n_j(t)=|\mathcal{W}_j(t)|$. Committee nodes operate the gateway,
authenticate transactions, construct cross-chain blocks, verify worker
responses, and issue committee certificates. Worker nodes maintain coded
fragments of the cross-chain settlement state and execute distributed
validation tasks. Workers do not participate in native consensus and have no
independent authority to confirm cross-chain blocks.

Let $\mathcal{U}_j$ denote the account set associated with chain $j$. The
account sets are disjoint, and
\[
\mathcal{U}
=
\bigcup_{j\in\mathcal{N}}\mathcal{U}_j,
\qquad
M
=
|\mathcal{U}|.
\]
The confirmed cross-chain settlement state at checkpoint $t$ is
$\mathbf{b}(t)\in\mathbb{Z}_{\geq 0}^{M}$. Committee
$\mathcal{C}_j(t)$ maintains the local checkpoint copy $\mathbf{b}_j(t)$.
For every active honest chain in interval $t$,
$\mathbf{b}_j(t)=\mathbf{b}(t)$.

During interval $t$, source chain $j$ forms a transfer batch represented
conceptually by
\[
\mathbf{X}_j(t)
\in
\mathbb{Z}_{\geq 0}^{M\times M},
\]
where $\mathbf{X}_j(t)[u,v]$ is the amount transferred from
$u\in\mathcal{U}_j$ to $v\in\mathcal{U}$. Entries associated with source
accounts outside $\mathcal{U}_j$ are zero. The matrix is a mathematical
representation only; an implementation stores the batch sparsely as transfer
records or as sparse debit and credit vectors. The batch induces
\begin{equation}
\mathbf{d}_j(t)
=
\mathbf{X}_j(t)\mathbf{1},
\qquad
\mathbf{c}_j(t)
=
\mathbf{X}_j^{\mathsf T}(t)\mathbf{1},
\label{eq:debit-credit}
\end{equation}
and the corresponding settlement-state increment is
\begin{equation}
\Delta\mathbf{b}_j(t)
=
\mathbf{c}_j(t)-\mathbf{d}_j(t).
\label{eq:balance-delta}
\end{equation}

Let $\mathcal{R}_j(t)$ denote the authenticated replay-protection state used
by chain $j$, including applied transaction identifiers, source-account
nonces, and accepted gateway sequence numbers. A batch $\mathbf{X}_j(t)$ is
admissible relative to $(\mathbf{b}_j(t),\mathcal{R}_j(t))$ if
\begin{equation}
\mathsf{Valid}\!\left(
\mathbf{X}_j(t)
\mid
\mathbf{b}_j(t),\mathcal{R}_j(t)
\right)
=
1,
\label{eq:batch-validity}
\end{equation}
which requires valid transaction signatures and native-finality certificates,
fresh transaction identifiers and nonces, the next expected issuer sequence
number, and
\begin{equation}
\mathbf{b}_j(t)-\mathbf{d}_j(t)
\succeq
\mathbf{0}.
\label{eq:balance-condition}
\end{equation}
The candidate post-batch state is
\begin{equation}
\mathbf{b}_j^{+}(t)
=
\mathbf{b}_j(t)+\Delta\mathbf{b}_j(t).
\label{eq:post-batch-state}
\end{equation}
Because
$\mathbf{1}^{\mathsf T}\mathbf{d}_j(t)
=
\mathbf{1}^{\mathsf T}\mathbf{c}_j(t)$,
each admissible transfer batch preserves the total amount represented in the
settlement layer.

The source gateway reserves the debit amount when the native lock event is
finalized. This reservation prevents the source accounts from reusing the
same funds while the batch is outstanding, but it is not applied as a second
debit to $\mathbf{b}(t)$. The settlement increment in
\eqref{eq:balance-delta} is applied exactly once, after cross-chain
confirmation. A batch that fails before DAG submission is aborted and its
source reservation is released. After DAG submission, the current model does
not permit timeout-based rollback; the reservation remains active until the
block confirms. Throughout this paper, each issuer has at most one
outstanding outgoing sequence. Supporting multiple pipelined sequences per
issuer would require a stronger reservation and conflict model and is outside
the present scope.

For coded storage, $\mathbf{b}_j(t)$ is zero-padded when necessary and
partitioned into $k_j(t)$ equal-length blocks,
\[
\mathbf{b}_j(t)
=
\left[
\mathbf{b}_{j,1}^{\mathsf T}(t),
\ldots,
\mathbf{b}_{j,k_j(t)}^{\mathsf T}(t)
\right]^{\mathsf T}.
\]
Define the power-of-two code length
\[
\bar n_j(t)
=
2^{\lceil\log_2 n_j(t)\rceil},
\]
and let
\[
\mathbf{G}_j(t)
\in
\mathbb{F}_Q^{\bar n_j(t)\times k_j(t)}
\]
be the reduced Polar generator matrix used during interval $t$. After
indexing physical workers by
$i\in\{1,\ldots,n_j(t)\}$, worker $i$ stores
\begin{equation}
\widetilde{\mathbf{b}}_{j,i}(t)
=
\sum_{\ell=1}^{k_j(t)}
\mathbf{G}_j(t)[i,\ell]\mathbf{b}_{j,\ell}(t),
\label{eq:persistent-coded-state}
\end{equation}
under the signed finite-field embedding defined in
Section~\ref{sec:polar-validation}. The remaining
$\bar n_j(t)-n_j(t)$ code positions are virtual erasures and store no data.
The worker assignment, information set, and generator matrix remain fixed
within a checkpoint interval and may change only at a confirmed
reconfiguration checkpoint.

When newly confirmed blocks induce the aggregate settlement increment
$\Delta\mathbf{b}_j^{\mathrm{conf}}(t)$, the committee encodes the
corresponding state-block increments and sends each physical worker only its
coded increment. Worker $i$ updates its persistent fragment according to
\begin{equation}
\widetilde{\mathbf{b}}_{j,i}(t+1)
=
\widetilde{\mathbf{b}}_{j,i}(t)
+
\Delta\widetilde{\mathbf{b}}_{j,i}^{\mathrm{conf}}(t).
\label{eq:coded-state-update}
\end{equation}
Thus, workers need not store the raw settlement state or reconstruct the
complete transfer history after every checkpoint.

The cross-chain ledger is a directed acyclic graph
$\mathcal{G}(t)=(\mathcal{V}(t),\mathcal{E}(t))$. A block issued by chain $j$
with gateway sequence number $r$ is denoted by $B_j(r)$. It contains a
natively finalized outgoing lock event, the associated transfer batch, a
coded-validation certificate, and a parent set. An edge
$(B',B)\in\mathcal{E}(t)$ means that the later block $B'$ directly approves
the earlier block $B$.

Let $\mathsf{Past}(B)$ and $\mathsf{Fut}_t(B)$ denote the past and future
cones of block $B$, respectively. Each participating chain $j$ is assigned a
protocol-level attestation weight $\omega_j>0$, where
\[
\sum_{j\in\mathcal{N}}\omega_j
=
1.
\]
These weights are membership parameters and are not obtained by directly
comparing native stake values across heterogeneous chains. The weight vector
and chain membership remain fixed within a checkpoint interval. Any
reconfiguration takes effect only at a subsequently confirmed checkpoint and
is not applied retroactively to blocks from an earlier interval.

A chain contributes its weight at most once to the support of a block. The
support set and aggregated weight of block $B$ in interval $t$ are
\begin{equation}
\begin{aligned}
\mathsf{Supp}_t(B)
&=
\left\{
j\in\mathcal{N}:
\exists\,B_j(r)
\in
\mathsf{Fut}_t(B)\cup\{B\}
\right\},\\
\mathsf{AW}_t(B)
&=
\sum_{j\in\mathsf{Supp}_t(B)}
\omega_j.
\end{aligned}
\label{eq:aggregated-weight}
\end{equation}
A block is confirmed when $\mathsf{AW}_t(B)$ reaches a threshold
$\eta\in(0,1)$.

Two blocks $B$ and $B'$ conflict, denoted by $B\perp B'$, if they cannot
both appear in a common valid settlement history. Conflicts include distinct
blocks with the same issuer and gateway sequence number, repeated transaction
identifiers or source-account nonces, and batches whose joint execution
violates~\eqref{eq:balance-condition}. Honest chains do not approve two
conflicting blocks.

\subsection{Adversarial Model and Design Objectives}
\label{subsec:adversarial-model}

Communication is partially synchronous: there exists an unknown global
stabilization time after which every message sent between honest participants
is delivered within a known bound $\Delta$.

The chain set is partitioned into honest and Byzantine subsets
$\mathcal{N}_{\mathrm{h}}$ and $\mathcal{N}_{\mathrm{a}}$. A Byzantine chain
may issue malformed or conflicting blocks, equivocate across DAG branches,
withhold attestations, select parents adversarially, or issue blocks at an
elevated rate. Its total attestation weight is bounded by
\begin{equation}
\sum_{j\in\mathcal{N}_{\mathrm{a}}}
\omega_j
\leq
\rho.
\label{eq:adversarial-weight}
\end{equation}
The confirmation threshold is chosen so that conflicting blocks cannot both
collect sufficient distinct-chain support.

For committee $\mathcal{C}_j(t)$, let
$n_j^{\mathrm{com}}(t)=|\mathcal{C}_j(t)|$, let
$f_j^{\mathrm{com}}(t)$ be the maximum number of Byzantine committee
members, and let $q_j(t)$ be the certificate threshold. We require
\begin{equation}
2q_j(t)
>
n_j^{\mathrm{com}}(t)+f_j^{\mathrm{com}}(t),
\qquad
q_j(t)
\leq
n_j^{\mathrm{com}}(t)-f_j^{\mathrm{com}}(t).
\label{eq:committee-threshold-condition}
\end{equation}
The first inequality ensures that two threshold quorums intersect in at
least one honest committee member, while the second permits an all-honest
quorum to make progress. The standard choice
\[
n_j^{\mathrm{com}}(t)
\geq
3f_j^{\mathrm{com}}(t)+1,
\qquad
q_j(t)
=
2f_j^{\mathrm{com}}(t)+1
\]
satisfies these conditions. Honest committee members sign at most one event
for a given chain, interval, stage, and sequence number, and only after the
prescribed local checks succeed. Byzantine committee members may reject valid
worker responses, propose inconsistent events, or delay progress, but they
cannot forge threshold certificates or native-finality certificates.

Workers may be honest, unresponsive, or Byzantine. An honest worker returns
the correct coded result. An unresponsive worker returns no accepted result
before the task deadline. A Byzantine worker may return an arbitrary value.
For a validation instance on chain $j$, let
$s_j^{\mathrm{wrk}}(t)$ and $f_j^{\mathrm{wrk}}(t)$ denote the numbers of
unresponsive and Byzantine workers, respectively. Polar coding provides
resilience to unavailable responses, whereas randomized verification detects
incorrect responses before decoding; rejected responses are treated as
erasures. Committee-held verification secrets are generated independently,
stored or secret-shared securely, and hidden from a worker until after that
worker commits to its response. Adaptive corruption that reveals these
verification vectors before commitment is outside the threat model.

Digital signatures and finality certificates are assumed unforgeable, hash
functions are collision resistant, and authenticated messages cannot be
modified without detection. The base protocol provides neither transaction
confidentiality nor privacy of the settlement state.

SPID is required to satisfy four properties. First, every confirmed block
must be valid relative to the confirmed state and deterministic checkpoint
order that precede it, and no two conflicting blocks may both become
confirmed. Second, apart from separately authenticated deposits and
withdrawals, the confirmed numerical state must satisfy
\begin{equation}
\mathbf{b}(t)
\succeq
\mathbf{0},
\qquad
\mathbf{1}^{\mathsf T}\mathbf{b}(t)
=
\mathbf{1}^{\mathsf T}\mathbf{b}(0).
\label{eq:state-conservation}
\end{equation}
Third, the persistent coded state maintained by every honest physical worker
must remain consistent with the confirmed raw state; specifically,
\begin{equation}
\widetilde{\mathbf{b}}_{j,i}(t)
=
\sum_{\ell=1}^{k_j(t)}
\mathbf{G}_j(t)[i,\ell]\mathbf{b}_{j,\ell}(t)
\label{eq:coded-state-invariant}
\end{equation}
must hold except with the stated cryptographic and randomized-verification
error. Fourth, after global stabilization, every admissible batch issued by
an honest chain must eventually be confirmed provided that the honest
committee can form certificates, the accepted worker-response pattern is
decodable, and honest chains continue issuing approvals with sufficient
aggregate weight.

For a cross-chain block $B$, let $T_{\mathrm{lock}}(B)$ be the time at which
the source gateway reserves the corresponding debit and let
$T_{\mathrm{release}}(B)$ be the time at which all destination credits
carried by $B$ have been applied. The end-to-end settlement latency is
\begin{equation}
T_{\mathrm{fin}}(B)
=
T_{\mathrm{release}}(B)
-
T_{\mathrm{lock}}(B).
\label{eq:settlement-latency}
\end{equation}
The objective is to reduce this latency under heterogeneous worker execution
and adversarial DAG activity while preserving safety, conservation,
coded-state consistency, and liveness.

\section{SPID Architecture and Protocol}
\label{sec:spid-protocol}

\subsection{Event-Certified Coded Validation}
\label{subsec:event-coded-validation}

SPID processes each outgoing transfer instance through four ordered stages:
\[
\mathsf{Prepare}
\longrightarrow
\mathsf{Validate}
\longrightarrow
\mathsf{Attest}
\longrightarrow
\mathsf{Commit}.
\]
Each transition is authorized by a committee-certified event. These events
provide transcript commitment, deterministic stage ordering, and replay
protection; they coordinate cross-chain execution but do not replace the
native consensus protocol of any participating chain.

Consider chain $j$ during checkpoint interval $t$, and let $r_j(t)$ be the
gateway sequence number of its active outgoing instance. For stage
$\ell\in\{0,1,2,3\}$, define
\begin{equation}
\varepsilon_{j,\ell}(t)
=
\Bigl(
j,
t,
r_j(t),
\ell,
h_{j,\ell-1}(t),
\alpha_{j,\ell}(t),
\beta_{j,\ell}(t),
\sigma_{j,\ell}(t)
\Bigr),
\label{eq:certified-event}
\end{equation}
where $h_{j,\ell-1}(t)$ is the hash of the preceding certified event,
$\alpha_{j,\ell}(t)$ and $\beta_{j,\ell}(t)$ commit to the stage input and
output, respectively, and $\sigma_{j,\ell}(t)$ is a valid threshold
certificate from $\mathcal{C}_j(t)$. For $\ell=0$, the predecessor value is
the confirmed checkpoint commitment $h^{\mathrm{cp}}(t)$.

A gateway accepts $\varepsilon_{j,\ell}(t)$ only if the chain identifier,
checkpoint interval, gateway sequence number, and stage index match the
expected local state; the predecessor hash is correct; the committee
certificate is valid; and the event identifier has not previously been
consumed. An honest committee member signs at most one event for a given
tuple $(j,t,r_j(t),\ell)$. Consequently, replaying, reordering, or replacing
a certified event invalidates every subsequent event in the same transcript.

In the $\mathsf{Prepare}$ stage, the source gateway verifies the transaction
signatures, source-account nonces, transaction identifiers, and expected
gateway sequence number of the proposed batch $\mathbf{X}_j(t)$. It then
reserves the debit vector $\mathbf{d}_j(t)$ under the native-chain execution
rules. Let $\zeta_j^{\mathrm{lock}}(t)$ denote a commitment to this reservation
record, including the batch hash, debit amounts, source accounts, and gateway
sequence number. The natively finalized outgoing event is
\begin{equation}
e_j(t)
=
\Bigl(
j,
t,
r_j(t),
h^{\mathrm{cp}}(t),
H\bigl(\mathbf{X}_j(t)\bigr),
\zeta_j^{\mathrm{lock}}(t),
\sigma_j^{\mathrm{nat}}(t)
\Bigr),
\label{eq:outgoing-lock-event}
\end{equation}
where $\sigma_j^{\mathrm{nat}}(t)$ is the corresponding native-finality
certificate. Native finality establishes that the reservation exists; it
does not constitute cross-chain confirmation.

The preparation event commits to the proposed batch, the finalized lock
event, and the replay-protection state against which the batch was formed:
\begin{equation}
\varepsilon_{j,0}(t)
=
\Bigl(
j,
t,
r_j(t),
0,
h^{\mathrm{cp}}(t),
H\bigl(\mathbf{X}_j(t),e_j(t)\bigr),
H\bigl(\mathcal{R}_j(t)\bigr),
\sigma_{j,0}(t)
\Bigr).
\label{eq:prepare-event}
\end{equation}

The committee next constructs the candidate-parent set from the public tips
visible in its local DAG view. A tip is preliminarily admissible if its
native-finality and validation certificates are well formed, its issuer and
gateway sequence number are valid, and its past cone does not conflict with
the confirmed history represented by $h^{\mathrm{cp}}(t)$.

To prevent discretionary selection, the committee derives the public seed
\begin{equation}
s_j(t)
=
H\Bigl(
h^{\mathrm{cp}}(t),
j,
r_j(t),
H\bigl(e_j(t)\bigr)
\Bigr).
\label{eq:parent-selection-seed}
\end{equation}
The seed determines a canonical pseudorandom permutation of the preliminarily
admissible tips. Scanning this order, the committee selects at most one block
from each issuer, excludes blocks issued by chain $j$, and retains at most
$K$ candidates. The resulting ordered candidate set is
\begin{equation}
\mathcal{Q}_j(t)
=
\left(
Q_{j,1}(t),
\ldots,
Q_{j,K_j(t)}(t)
\right),
\qquad
K_j(t)\leq K.
\label{eq:candidate-parent-set}
\end{equation}
The analysis in Section~\ref{sec:security} models the permutation induced by
$s_j(t)$ as conditionally uniform over the admissible public tips.

The outgoing batch is appended to the candidate sequence, producing the
joint validation workload
\begin{equation}
\mathcal{V}_j(t)
=
\left(
Q_{j,1}(t),
\ldots,
Q_{j,K_j(t)}(t),
\mathbf{X}_j(t)
\right),
\qquad
L_j(t)
=
K_j(t)+1.
\label{eq:joint-validation-set}
\end{equation}
For $\ell\in\{1,\ldots,L_j(t)\}$, let
$\mathbf{d}_{j,\ell}(t)$ and $\mathbf{c}_{j,\ell}(t)$ denote the debit and
credit vectors of the $\ell$th workload item. The final item corresponds to
the local batch:
\[
\mathbf{d}_{j,L_j(t)}(t)
=
\mathbf{d}_j(t),
\qquad
\mathbf{c}_{j,L_j(t)}(t)
=
\mathbf{c}_j(t).
\]

The candidate-selection rule permits at most one unconfirmed block from each
issuer and excludes chain $j$ from its own candidate-parent set. Because only
an issuer may debit accounts in its own account set, the debit supports of
distinct workload items are disjoint:
\begin{equation}
\operatorname{supp}
\bigl(
\mathbf{d}_{j,\ell}(t)
\bigr)
\cap
\operatorname{supp}
\bigl(
\mathbf{d}_{j,\ell'}(t)
\bigr)
=
\varnothing,
\qquad
\ell\neq\ell'.
\label{eq:disjoint-debit-support}
\end{equation}
Consequently, each sufficient-funds condition can be evaluated against the
same confirmed checkpoint state:
\begin{equation}
\mathbf{b}_j(t)
-
\mathbf{d}_{j,\ell}(t)
\succeq
\mathbf{0}.
\label{eq:joint-candidate-balance-check}
\end{equation}
Credits created by one workload item are not treated as spendable by another
item in the same validation instance.

The committee additionally verifies the workload as a joint object. It
checks that transaction identifiers, nonces, and issuer sequence numbers are
fresh relative to $\mathcal{R}_j(t)$ and mutually distinct across the
workload; that all required signatures and certificates are valid; and that
the union of the candidate past cones is conflict-free. These authenticated
checks are performed separately from the coded numerical computation.

The numerical checks in~\eqref{eq:joint-candidate-balance-check} are evaluated
from the persistent coded state. The committee constructs a query description
$\mathsf{Qry}_j(t)$ specifying the required state coordinates, encoded debit
vectors, workload ordering, field parameters, and response deadline. It also
constructs a worker assignment $\mathsf{Asn}_j(t)$. The validation-trigger
event is
\begin{equation}
\varepsilon_{j,1}(t)
=
\Bigl(
j,
t,
r_j(t),
1,
H\bigl(\varepsilon_{j,0}(t)\bigr),
H\bigl(
\mathcal{V}_j(t),
\mathsf{Qry}_j(t),
\mathbf{G}_j(t)
\bigr),
H\bigl(\mathsf{Asn}_j(t)\bigr),
\sigma_{j,1}(t)
\Bigr).
\label{eq:validation-trigger-event}
\end{equation}

Each physical worker receives one task bundle derived from
$\mathsf{Qry}_j(t)$ and returns one committed, signed batched response
containing a coded result for every item in $\mathcal{V}_j(t)$. The committee
verifies each returned row using the secret linear checks developed in
Section~\ref{sec:polar-validation}. A missing response, an invalid signature,
a commitment mismatch, or a response that fails a linear check is treated as
an erasure.

Let $\mathcal{A}_j^{\mathrm{rsp}}(t)$ denote the indices of the accepted
physical-worker responses. Once this response pattern is decodable for the
selected generator matrix and information set, the committee reconstructs
the required post-debit state values and evaluates
\eqref{eq:joint-candidate-balance-check}. It combines the reconstructed
numerical results with the signature, replay, sequence-number, certificate,
and DAG-conflict checks.

Let
\[
\mathcal{Q}_j^{\mathrm{val}}(t)
\subseteq
\mathcal{Q}_j(t)
\]
denote the candidate-parent blocks that pass all checks, and let
$\xi_j(t)\in\{0,1\}$ indicate whether the local outgoing batch passes all
checks. The committee produces the validation certificate
\begin{equation}
\pi_j(t)
=
\mathsf{Cert}_{\mathcal{C}_j(t)}
\left(
\begin{array}{c}
j,\ t,\ r_j(t),\ h^{\mathrm{cp}}(t),\\
H\bigl(\varepsilon_{j,0:1}(t)\bigr),\
H\bigl(\mathcal{Q}_j^{\mathrm{val}}(t)\bigr),\\
\xi_j(t),\
H\bigl(\mathcal{T}_j^{\mathrm{wrk}}(t)\bigr)
\end{array}
\right),
\label{eq:validation-certificate}
\end{equation}
where $\mathcal{T}_j^{\mathrm{wrk}}(t)$ is the accepted worker transcript and
$\varepsilon_{j,0:1}(t)$ denotes the ordered pair of certified events
$\varepsilon_{j,0}(t)$ and $\varepsilon_{j,1}(t)$.

The $\mathsf{Attest}$ event commits to the validation result:
\begin{equation}
\varepsilon_{j,2}(t)
=
\Bigl(
j,
t,
r_j(t),
2,
H\bigl(\varepsilon_{j,1}(t)\bigr),
H\bigl(
\mathcal{Q}_j^{\mathrm{val}}(t),
\xi_j(t)
\bigr),
H\bigl(\pi_j(t)\bigr),
\sigma_{j,2}(t)
\Bigr).
\label{eq:attestation-event}
\end{equation}

If $\xi_j(t)=0$, the outgoing instance is aborted before DAG submission and
the source reservation is released. If $\xi_j(t)=1$, the committee proceeds
to deterministic parent selection and block construction.

\subsection{DAG Attachment and Cross-Chain Settlement}
\label{subsec:dag-settlement}

The parent set of the local block is selected from
$\mathcal{Q}_j^{\mathrm{val}}(t)$ using the canonical order induced by
$s_j(t)$. The committee scans the validated candidates in that order. A
candidate is retained only if the union of its past cone, the past cones of
the already retained candidates, and the local outgoing batch remains
conflict-free. The resulting parent set satisfies
\begin{equation}
\mathcal{P}_j(t)
\subseteq
\mathcal{Q}_j^{\mathrm{val}}(t),
\qquad
|\mathcal{P}_j(t)|
\leq
K.
\label{eq:parent-set}
\end{equation}
Because the selection rule and its seed are public, another participant can
reproduce the scan and verify that the parent set was not chosen
discretionarily.

Chain $j$ constructs the cross-chain block
\begin{equation}
B_j\bigl(r_j(t)\bigr)
=
\Bigl(
j,
t,
r_j(t),
h^{\mathrm{cp}}(t),
e_j(t),
\mathbf{X}_j(t),
\pi_j(t),
\mathcal{P}_j(t),
H\bigl(\varepsilon_{j,0:2}(t)\bigr)
\Bigr),
\label{eq:spid-block}
\end{equation}
where $\varepsilon_{j,0:2}(t)$ denotes the ordered event transcript through
the $\mathsf{Attest}$ stage. The block directly approves every block in
$\mathcal{P}_j(t)$ and indirectly approves the blocks in their past cones.

Before inserting a received block into its local DAG view, an honest
participant verifies the native-finality certificate and source-lock
commitment; the issuer identity and gateway sequence number; the referenced
confirmed checkpoint; the validation certificate and worker-transcript
commitment; the certified-event transcript; the transaction identifiers,
nonces, and replay metadata; and the reproducibility and conflict-freedom of
the parent set.

A block that fails any check is rejected and contributes no approval weight.
An accepted block issued by chain $i$ contributes $\omega_i$ to every block
in its past cone, but contributes that weight at most once to any supported
block. As defined in Section~\ref{sec:system-model}, block $B$ becomes
confirmed when
\begin{equation}
\mathsf{AW}_t(B)
\geq
\eta.
\label{eq:block-confirmation}
\end{equation}

A confirmation certificate $\kappa(B)$ identifies a set of descendant blocks,
their distinct issuers, and the corresponding DAG inclusion proofs. It is
valid if each descendant is valid, each inclusion proof establishes support
for $B$, no issuer contributes more than once, and the total issuer weight is
at least $\eta$.

Let $\mathcal{F}_j(t)$ be the set of blocks newly confirmed by chain $j$
while advancing from checkpoint $t$ to checkpoint $t+1$. Confirmed blocks
are applied according to a deterministic total order
$\prec_{\mathrm{cp}}$ that extends the DAG partial order. Blocks that are
otherwise incomparable are ordered by a fixed tie-breaking rule over their
issuer identifiers, gateway sequence numbers, and block hashes.

The aggregate numerical update is
\begin{equation}
\Delta\mathbf{b}_j^{\mathrm{conf}}(t)
=
\sum_{B\in\mathcal{F}_j(t)}
\left(
\mathbf{c}_B-\mathbf{d}_B
\right),
\label{eq:confirmed-balance-update}
\end{equation}
where the summation is evaluated according to $\prec_{\mathrm{cp}}$. The raw
checkpoint state evolves as
\begin{equation}
\mathbf{b}_j(t+1)
=
\mathbf{b}_j(t)
+
\Delta\mathbf{b}_j^{\mathrm{conf}}(t).
\label{eq:raw-state-evolution}
\end{equation}
The replay-protection state $\mathcal{R}_j(t+1)$ is updated in the same order
with the confirmed transaction identifiers, nonces, and issuer sequence
numbers.

The committee partitions
$\Delta\mathbf{b}_j^{\mathrm{conf}}(t)$ using the same state-block layout as
$\mathbf{b}_j(t)$ and encodes it with $\mathbf{G}_j(t)$. Worker $i$ receives
only its coded increment
$\Delta\widetilde{\mathbf{b}}_{j,i}^{\mathrm{conf}}(t)$ and applies
\eqref{eq:coded-state-update}. By linearity,
\begin{align}
\widetilde{\mathbf{b}}_{j,i}(t+1)
&=
\widetilde{\mathbf{b}}_{j,i}(t)
+
\Delta\widetilde{\mathbf{b}}_{j,i}^{\mathrm{conf}}(t)
\nonumber\\
&=
\sum_{\ell=1}^{k_j(t)}
\mathbf{G}_j(t)[i,\ell]
\left(
\mathbf{b}_{j,\ell}(t)
+
\Delta\mathbf{b}_{j,\ell}^{\mathrm{conf}}(t)
\right)
\nonumber\\
&=
\sum_{\ell=1}^{k_j(t)}
\mathbf{G}_j(t)[i,\ell]
\mathbf{b}_{j,\ell}(t+1).
\label{eq:coded-update-linearity}
\end{align}
Thus, the coded-state invariant is preserved without redistributing the
complete raw state.

The checkpoint update, replay-protection root, confirmation certificates, and
coded-increment commitments are recorded in the $\mathsf{Commit}$ event:
\begin{equation}
\varepsilon_{j,3}(t)
=
\Bigl(
j,
t,
r_j(t),
3,
H\bigl(\varepsilon_{j,2}(t)\bigr),
H\bigl(
\mathcal{F}_j(t),
\kappa(\mathcal{F}_j(t))
\bigr),
H\bigl(
\mathbf{b}_j(t+1),
\mathcal{R}_j(t+1),
\Delta\widetilde{\mathbf{S}}_j^{\mathrm{conf}}(t)
\bigr),
\sigma_{j,3}(t)
\Bigr).
\label{eq:commit-event}
\end{equation}
Here, $\kappa(\mathcal{F}_j(t))$ denotes the collection of confirmation
certificates for the newly confirmed blocks. The commitment to
$\Delta\widetilde{\mathbf{S}}_j^{\mathrm{conf}}(t)$ may be implemented as a
Merkle root or another collision-resistant vector commitment.

For a confirmed block $B_i(r)$, destination chain $d$ extracts the transfers
addressed to accounts in $\mathcal{U}_d$. Its gateway verifies the
confirmation certificate, originating native-finality certificate,
validation certificate, and transaction identifiers. The gateway then
atomically records each fresh transaction identifier and applies the
corresponding destination credit. A transaction whose identifier is already
present in $\mathcal{R}_d$ is ignored. This atomic record-and-apply rule
provides exactly-once destination execution.

After confirmation, the source gateway marks the corresponding reservation
as consumed. Consuming the native reservation does not apply an additional
debit to the cross-chain settlement state: the debit is already included
exactly once in~\eqref{eq:confirmed-balance-update}. Similarly, destination
application realizes the confirmed credit on the destination chain but does
not create a second settlement-state credit.

A batch may be aborted and its reservation released only before DAG
submission. Once $B_j(r_j(t))$ has been submitted, the reservation remains
active until the block is confirmed; the present protocol does not include a
timeout-based rollback rule. Under the liveness conditions established in
Section~\ref{sec:security}, a valid block issued by an honest chain eventually
collects sufficient distinct-chain support, its destination transfers are
applied exactly once, its source reservation is consumed, and its increment
is incorporated into both the raw and persistently coded checkpoint states.

\section{Verifiable Polar-Coded State Validation}
\label{sec:polar-validation}

\subsection{Persistent Polar-Coded State}
\label{subsec:persistent-polar-state}

Fix a chain $j$ and checkpoint interval $t$. Let
$n_j(t)=|\mathcal{W}_j(t)|$ be the number of physical workers and define the
power-of-two code length
\begin{equation}
\bar n_j(t)
=
2^{\lceil\log_2 n_j(t)\rceil}.
\label{eq:polar-code-length}
\end{equation}
The additional $\bar n_j(t)-n_j(t)$ codeword positions are virtual and are
modeled as permanent erasures. Let $k_j(t)<\bar n_j(t)$ be the number of
uncoded state blocks and
\begin{equation}
R_j^{\mathrm{code}}(t)
=
\frac{k_j(t)}{\bar n_j(t)}
\label{eq:polar-coding-rate}
\end{equation}
be the nominal coding rate.

Let
\[
m_j(t)
=
\left\lceil
\frac{M}{k_j(t)}
\right\rceil.
\]
After appending zero-valued padding coordinates, the confirmed state is
reshaped as
\begin{equation}
\mathbf{S}_j(t)
=
\begin{bmatrix}
\mathbf{b}_{j,1}^{\mathsf T}(t)\\
\vdots\\
\mathbf{b}_{j,k_j(t)}^{\mathsf T}(t)
\end{bmatrix}
\in
\mathbb{F}_Q^{k_j(t)\times m_j(t)}.
\label{eq:state-block-matrix}
\end{equation}
The padded coordinates remain zero and are excluded from every validity
check.

All coded operations use an odd-prime field $\mathbb{F}_Q$. Let
$A_{\max}$ be a protocol-level upper bound on the absolute value of every
uncoded scalar that may be decoded within one checkpoint interval, including
balances, debits, post-debit values, and accumulated checkpoint increments.
We require
\begin{equation}
Q
>
2A_{\max}.
\label{eq:field-size-condition}
\end{equation}
An integer $a\in[-A_{\max},A_{\max}]$ is embedded as $a\bmod Q$. After
decoding, the canonical representative in
$[-(Q-1)/2,(Q-1)/2]$ is selected. Condition~\eqref{eq:field-size-condition}
therefore makes the recovered integer unique, even though intermediate coded
operations are performed modulo $Q$.

Define the Polar kernel over $\mathbb{F}_Q$ by
\[
\mathbf{K}_2
=
\begin{bmatrix}
1&0\\
1&1
\end{bmatrix},
\]
and let
\begin{equation}
\mathbf{T}_{\bar n_j(t)}
=
\mathbf{P}_{\bar n_j(t)}
\mathbf{K}_2^{\otimes\log_2\bar n_j(t)},
\label{eq:polar-transform}
\end{equation}
where $\mathbf{P}_{\bar n_j(t)}$ is the bit-reversal permutation matrix.
The information set
$\mathcal{I}_j(t)\subseteq\{1,\ldots,\bar n_j(t)\}$ has cardinality
$k_j(t)$. With the uncoded state blocks represented as columns of the
encoder input, the reduced generator matrix is
\begin{equation}
\mathbf{G}_j(t)
=
\mathbf{T}_{\bar n_j(t)}
\bigl[\mathcal{I}_j(t),:\bigr]^{\mathsf T}
\in
\mathbb{F}_Q^{\bar n_j(t)\times k_j(t)}.
\label{eq:reduced-polar-generator}
\end{equation}
This orientation is equivalent to placing the $k_j(t)$ uncoded blocks in the
information positions, setting the frozen positions to zero, and applying
the transposed Polar transform.

The coded state is
\begin{equation}
\widetilde{\mathbf{S}}_j(t)
=
\mathbf{G}_j(t)\mathbf{S}_j(t)
\in
\mathbb{F}_Q^{\bar n_j(t)\times m_j(t)}.
\label{eq:polar-state-encoding}
\end{equation}
After absorbing the reliability-aware worker-to-position assignment into the
indexing, physical worker $i\in\{1,\ldots,n_j(t)\}$ stores row $i$:
\[
\widetilde{\mathbf{b}}_{j,i}^{\mathsf T}(t)
=
\widetilde{\mathbf{S}}_j(t)[i,:].
\]
No data are assigned to the virtual positions.

Let $p_{j,i}(t)$ be the probability that codeword position $i$ produces no
accepted response before the validation deadline. For a virtual position,
$p_{j,i}(t)=1$. Treating these positions as independent erasure channels, the
synthesized erasure parameters are obtained recursively from
\begin{equation}
z^{-}
=
z_1+z_2-z_1z_2,
\qquad
z^{+}
=
z_1z_2.
\label{eq:polar-reliability-recursion}
\end{equation}
Let $Z_{j,r}(t)$ be the resulting erasure parameter of synthesized channel
$r$. The information set contains the $k_j(t)$ smallest values of
$Z_{j,r}(t)$, subject to
\begin{equation}
\sum_{r\in\mathcal{I}_j(t)}
Z_{j,r}(t)
\leq
\epsilon_{\mathrm{dec}}.
\label{eq:information-set-condition}
\end{equation}
This construction criterion controls the decoding-failure probability under
the independent-erasure model; algebraic recoverability for a realized
response pattern is characterized separately below.

Suppose no worker or code reconfiguration occurs at checkpoint $t+1$. Let
$\Delta\mathbf{S}_j^{\mathrm{conf}}(t)$ be the block-matrix representation of
the confirmed state increment. The committee computes
\begin{equation}
\Delta\widetilde{\mathbf{S}}_j^{\mathrm{conf}}(t)
=
\mathbf{G}_j(t)
\Delta\mathbf{S}_j^{\mathrm{conf}}(t)
\label{eq:encoded-confirmed-increment}
\end{equation}
and sends physical worker $i$ only row $i$. The worker applies
\begin{equation}
\widetilde{\mathbf{b}}_{j,i}^{\mathsf T}(t+1)
=
\widetilde{\mathbf{b}}_{j,i}^{\mathsf T}(t)
+
\Delta\widetilde{\mathbf{S}}_j^{\mathrm{conf}}(t)[i,:].
\label{eq:worker-persistent-update}
\end{equation}
Consequently,
\begin{equation}
\widetilde{\mathbf{S}}_j(t+1)
=
\mathbf{G}_j(t)
\left(
\mathbf{S}_j(t)
+
\Delta\mathbf{S}_j^{\mathrm{conf}}(t)
\right)
=
\mathbf{G}_j(t)\mathbf{S}_j(t+1).
\label{eq:persistent-state-invariance}
\end{equation}

If the worker set, block partition, or information set changes, the change is
activated only at a confirmed reconfiguration checkpoint. The committee then
re-encodes the current raw state under $\mathbf{G}_j(t+1)$ and refreshes the
verification material described next. Incremental updates resume after the
new configuration becomes active.

\subsection{Joint Validation and Verifiable Recovery}
\label{subsec:joint-coded-validation}

During interval $t$, chain $j$ validates the workload
$\mathcal{V}_j(t)$ from~\eqref{eq:joint-validation-set}, containing
$L_j(t)=K_j(t)+1$ items. For workload item $\ell$, let
\[
\mathbf{D}_{j,\ell}(t)
\in
\mathbb{F}_Q^{k_j(t)\times m_j(t)}
\]
be the block-matrix representation of its debit vector
$\mathbf{d}_{j,\ell}(t)$. The committee encodes
\begin{equation}
\widetilde{\mathbf{D}}_{j,\ell}(t)
=
\mathbf{G}_j(t)\mathbf{D}_{j,\ell}(t),
\qquad
\ell=1,\ldots,L_j(t).
\label{eq:encoded-candidate-debit}
\end{equation}

Physical worker $i$ receives row $i$ of every encoded debit matrix and
computes
\begin{equation}
\widetilde{\mathbf{y}}_{j,i,\ell}^{\mathsf T}(t)
=
\widetilde{\mathbf{b}}_{j,i}^{\mathsf T}(t)
-
\widetilde{\mathbf{D}}_{j,\ell}(t)[i,:],
\qquad
\ell=1,\ldots,L_j(t).
\label{eq:worker-candidate-result}
\end{equation}
The worker returns one signed, committed response bundle
\begin{equation}
\widetilde{\mathbf{Y}}_{j,i}(t)
=
\begin{bmatrix}
\widetilde{\mathbf{y}}_{j,i,1}^{\mathsf T}(t)\\
\vdots\\
\widetilde{\mathbf{y}}_{j,i,L_j(t)}^{\mathsf T}(t)
\end{bmatrix}
\in
\mathbb{F}_Q^{L_j(t)\times m_j(t)}.
\label{eq:batched-worker-response}
\end{equation}
For each $\ell$, the collection of worker rows is a Polar encoding of
$\mathbf{S}_j(t)-\mathbf{D}_{j,\ell}(t)$.

To detect incorrect coded results, the committee maintains hidden linear
verification material for each physical worker. Let $\gamma_j(t)$ be the
number of independent checks. For worker $i$, the committee samples
\[
\mathbf{r}_{j,i,1}(t),\ldots,
\mathbf{r}_{j,i,\gamma_j(t)}(t)
\overset{\$}{\leftarrow}
\mathbb{F}_Q^{m_j(t)}
\]
and stores
\begin{equation}
\tau_{j,i,h}(t)
=
\left\langle
\mathbf{r}_{j,i,h}(t),
\widetilde{\mathbf{b}}_{j,i}(t)
\right\rangle,
\qquad
h=1,\ldots,\gamma_j(t).
\label{eq:persistent-state-tags}
\end{equation}
The verification vectors and tags may be secret-shared among committee
members under a threshold that preserves confidentiality against up to
$f_j^{\mathrm{com}}(t)$ corruptions. Workers do not learn the vectors. The
soundness bound below is per validation instance and assumes that a worker
commits before obtaining information about the verification vectors used for
that instance. In adaptive deployments, these vectors should be generated
fresh, or derived from a domain-separated secret seed, for each validation
instance; long-lived vectors are appropriate only when acceptance outcomes do
not leak usable information before subsequent commitments.

For an ordinary checkpoint update, the tags evolve with the coded state:
\begin{equation}
\tau_{j,i,h}(t+1)
=
\tau_{j,i,h}(t)
+
\left\langle
\mathbf{r}_{j,i,h}(t),
\Delta\widetilde{\mathbf{S}}_j^{\mathrm{conf}}(t)[i,:]
\right\rangle.
\label{eq:tag-update}
\end{equation}
Thus, the committee need not recompute the tags from the complete settlement
history for verification vectors that remain active during a configuration
epoch.

After worker $i$ commits to its response bundle, the committee verifies, for
each workload item $\ell$ and check $h$,
\begin{equation}
\left\langle
\mathbf{r}_{j,i,h}(t),
\widetilde{\mathbf{y}}_{j,i,\ell}(t)
\right\rangle
=
\tau_{j,i,h}(t)
-
\left\langle
\mathbf{r}_{j,i,h}(t),
\widetilde{\mathbf{D}}_{j,\ell}(t)[i,:]
\right\rangle.
\label{eq:worker-result-verification}
\end{equation}
A worker index is accepted for the validation instance only if its signature,
commitment, dimensions, and every check in
\eqref{eq:worker-result-verification} are valid for all $L_j(t)$ workload
items. Otherwise, the entire response bundle is treated as an erasure. This
all-or-nothing rule yields one common accepted response set for the complete
joint workload.

If a Byzantine worker commits to an incorrect row
\[
\widehat{\mathbf{y}}_{j,i,\ell}(t)
\neq
\widetilde{\mathbf{y}}_{j,i,\ell}(t)
\]
before obtaining information about the secret verification vectors, then
\begin{equation}
\Pr\!\left[
\widehat{\mathbf{y}}_{j,i,\ell}(t)
\text{ passes every check}
\right]
\leq
Q^{-\gamma_j(t)}.
\label{eq:single-worker-soundness}
\end{equation}
Accordingly, if at most $f_j^{\mathrm{wrk}}(t)$ workers return incorrect
bundles, the per-instance false-acceptance probability satisfies
\begin{equation}
\epsilon_{\mathrm{ver},j}(t)
\leq
f_j^{\mathrm{wrk}}(t)L_j(t)Q^{-\gamma_j(t)}
+
\epsilon_{\mathrm{auth},j}(t),
\label{eq:validation-soundness}
\end{equation}
where $\epsilon_{\mathrm{auth},j}(t)$ accounts once for failures of the
signature, commitment, secret-sharing, and committee-certification
mechanisms.

Let
\[
\mathcal{A}_j^{\mathrm{rsp}}(t)
\subseteq
\{1,\ldots,n_j(t)\}
\]
be the accepted physical-worker indices, and let
$\mathbf{G}_{j,\mathcal{A}}(t)$ denote the submatrix of
$\mathbf{G}_j(t)$ formed by the corresponding rows. The realized response
pattern is algebraically decodable if
\begin{equation}
\operatorname{rank}
\left(
\mathbf{G}_{j,\mathcal{A}_j^{\mathrm{rsp}}(t)}(t)
\right)
=
k_j(t).
\label{eq:polar-decodability}
\end{equation}
The virtual positions never belong to
$\mathcal{A}_j^{\mathrm{rsp}}(t)$.

The implementation first applies successive-cancellation erasure decoding,
which exploits the Polar transform and has near-linear complexity. If that
decoder stops on a pattern that nevertheless satisfies
\eqref{eq:polar-decodability}, the committee applies a rank-based fallback
decoder. Thus, the information set is designed using Polar erasure
reliabilities, whereas the final algebraic recovery condition is full column
rank. The exact recovery-time analysis in Section~\ref{sec:security} uses the
rank condition rather than assuming that every set of $k_j(t)$ responses is
sufficient.

For each workload item $\ell$, successful recovery yields
\begin{equation}
\mathbf{Y}_{j,\ell}(t)
=
\mathbf{S}_j(t)-\mathbf{D}_{j,\ell}(t).
\label{eq:decoded-post-debit-state}
\end{equation}
The committee maps the recovered field elements to their canonical signed
integer representatives and ignores the padded coordinates. Item $\ell$
passes its numerical test if every remaining entry of
$\mathbf{Y}_{j,\ell}(t)$ is nonnegative. These results are combined with the
signature, certificate, replay, sequence-number, and DAG-conflict checks to
produce $\mathcal{Q}_j^{\mathrm{val}}(t)$ and $\xi_j(t)$ in
\eqref{eq:validation-certificate}.

Under the independent-erasure model used to construct the information set,
let $P_{\mathrm{rank},j}(t)$ and $P_{\mathrm{SC},j}(t)$ denote the failure
probabilities of rank recovery and successive-cancellation decoding,
respectively. Because the rank fallback can recover every full-column-rank
pattern,
\begin{equation}
P_{\mathrm{rank},j}(t)
\leq
P_{\mathrm{SC},j}(t)
\leq
\sum_{r\in\mathcal{I}_j(t)}
Z_{j,r}(t)
\leq
\epsilon_{\mathrm{dec}}.
\label{eq:polar-failure-bound}
\end{equation}
This probabilistic bound concerns unavailable or rejected responses. The
separate bound in~\eqref{eq:validation-soundness} concerns an incorrect
response that evades verification.

Persistent storage avoids re-encoding the complete state for every
validation instance. In a dense worst-case representation, encoding the
$L_j(t)$ debit matrices requires
\[
O\!\left(
L_j(t)\bar n_j(t)m_j(t)\log\bar n_j(t)
\right)
\]
field operations using the fast Polar transform. Sparse debit vectors reduce
this cost in proportion to the number of touched state coordinates. Each
worker performs $O(L_j(t)m_j(t))$ field operations and returns
$L_j(t)m_j(t)$ field symbols. Verification costs
\[
O\!\left(
\gamma_j(t)L_j(t)m_j(t)
\right)
\]
operations per responding worker. Successive-cancellation decoding costs
\[
O\!\left(
L_j(t)\bar n_j(t)m_j(t)\log\bar n_j(t)
\right).
\]
For an accepted set of size
$a_j(t)=|\mathcal{A}_j^{\mathrm{rsp}}(t)|$, a rank-based fallback requires
$O(a_j(t)k_j^2(t))$ operations to factor the response matrix and
\[
O\!\left(
L_j(t)a_j(t)k_j(t)m_j(t)
\right)
\]
operations to recover all workload items.

The parent limit $K$ and coding rate $R_j^{\mathrm{code}}(t)$ therefore
affect both layers of the protocol. Increasing $K$ permits a block to
approve more DAG branches but increases the joint validation workload.
Decreasing $R_j^{\mathrm{code}}(t)$ improves erasure tolerance but increases
storage, communication, and decoding cost. Their combined effect is
incorporated into the cross-layer analysis in Section~\ref{sec:security} and
evaluated in Section~\ref{sec:evaluation}.

\section{Cross-Layer Correctness and Performance Analysis}
\label{sec:security}

This section derives the analytical guarantees of SPID from the protocol
model in Sections~\ref{sec:system-model}--\ref{sec:polar-validation}.
The recovery, verification, and quorum results establish the local conditions
required by the cross-layer analysis. The principal result then identifies
how the coded-validation completion law determines the effective honest
issuance rate and, through it, the stability region of the cross-chain DAG.

\subsection{Coded Validation and Weighted Safety}
\label{subsec:coded-analysis}

Fix a chain $j$ and checkpoint interval $t$, and suppress these indices where
doing so causes no ambiguity. Write
\[
n=n_j(t),
\qquad
\bar n=\bar n_j(t),
\qquad
k=k_j(t),
\qquad
L=L_j(t),
\qquad
\gamma=\gamma_j(t),
\]
and let
$\mathbf{G}\in\mathbb{F}_Q^{\bar n\times k}$
be the reduced Polar generator matrix in
\eqref{eq:reduced-polar-generator}.

For $\mathcal{A}\subseteq[\bar n]$, where
$[\bar n]=\{1,\ldots,\bar n\}$, let
$\mathbf{G}_{\mathcal{A}}$ denote the row submatrix indexed by
$\mathcal{A}$. Define the algebraically decodable response-set family
\begin{equation}
\mathfrak{D}(\mathbf{G})
=
\left\{
\mathcal{A}\subseteq[\bar n]:
\operatorname{rank}
\left(
\mathbf{G}_{\mathcal{A}}
\right)
=
k
\right\}.
\label{eq:decodable-family}
\end{equation}
The family $\mathfrak{D}(\mathbf{G})$ is upward closed under set inclusion.

For codeword position $i$, let
$T_i^{(L)}\in[0,\infty]$ be the time at which the committee has received and
accepted the correct response bundle for all $L$ workload items. This time
includes worker execution, communication, signature and commitment
verification, and the linear checks in
\eqref{eq:worker-result-verification}. For a virtual position,
$T_i^{(L)}=\infty$ almost surely. Define
\begin{equation}
F_i^{(L)}(\tau)
=
\Pr\!\left[
T_i^{(L)}
\leq
\tau
\right],
\qquad
\mathcal{A}^{(L)}(\tau)
=
\left\{
i\in[\bar n]:
T_i^{(L)}
\leq
\tau
\right\}.
\label{eq:response-process}
\end{equation}
The rank-recovery time is
\begin{equation}
T_{\mathrm{rec}}^{(L)}
=
\inf
\left\{
\tau\geq0:
\mathcal{A}^{(L)}(\tau)
\in
\mathfrak{D}(\mathbf{G})
\right\}.
\label{eq:recovery-stopping-time}
\end{equation}

\begin{proposition}[Exact heterogeneous recovery law]
\label{prop:recovery-time-law}
Suppose that
$\{T_i^{(L)}\}_{i=1}^{\bar n}$
are mutually independent. Then, for every $\tau\geq0$,
\begin{equation}
\Pr\!\left[
T_{\mathrm{rec}}^{(L)}
\leq
\tau
\right]
=
\sum_{\mathcal{A}\in\mathfrak{D}(\mathbf{G})}
\prod_{i\in\mathcal{A}}
F_i^{(L)}(\tau)
\prod_{i\in[\bar n]\setminus\mathcal{A}}
\left(
1-F_i^{(L)}(\tau)
\right).
\label{eq:exact-recovery-cdf}
\end{equation}
If $\mathbb{E}[T_{\mathrm{rec}}^{(L)}]<\infty$, then
\begin{equation}
\mathbb{E}
\left[
T_{\mathrm{rec}}^{(L)}
\right]
=
\int_{0}^{\infty}
\Pr\!\left[
T_{\mathrm{rec}}^{(L)}
>
\tau
\right]
\mathrm{d}\tau.
\label{eq:exact-recovery-mean}
\end{equation}
\end{proposition}

The proof is given in Appendix~\ref{app:proof-recovery-law}. In contrast to
an MDS recovery law, \eqref{eq:exact-recovery-cdf} depends on the identities
of the responding positions through
$\mathfrak{D}(\mathbf{G})$, not only on their cardinality.

For identically distributed physical workers, define
\begin{equation}
a_r(\mathbf{G})
=
\left|
\left\{
\mathcal{A}\subseteq[n]:
|\mathcal{A}|=r,\
\operatorname{rank}
\left(
\mathbf{G}_{\mathcal{A}}
\right)
=
k
\right\}
\right|,
\qquad
r=k,\ldots,n.
\label{eq:decodability-spectrum}
\end{equation}

\begin{corollary}[Decodability-spectrum specialization]
\label{cor:recovery-spectrum}
Suppose that
$F_i^{(L)}=F^{(L)}$
for every $i\in[n]$ and
$F_i^{(L)}=0$
for every $i\in[\bar n]\setminus[n]$. Then
\begin{equation}
\Pr\!\left[
T_{\mathrm{rec}}^{(L)}
\leq
\tau
\right]
=
\sum_{r=k}^{n}
a_r(\mathbf{G})
\left(
F^{(L)}(\tau)
\right)^r
\left(
1-F^{(L)}(\tau)
\right)^{n-r}.
\label{eq:spectrum-recovery-cdf}
\end{equation}
\end{corollary}

We next quantify the residual error introduced by Byzantine worker
responses. For workload item $\ell$, let
$\mathbf{Y}_{\ell}$ denote the correct matrix in
\eqref{eq:decoded-post-debit-state}, and let
$\widehat{\mathbf{Y}}_{\ell}$ denote the matrix reconstructed by the
committee. Define
\begin{equation}
\mathcal{E}_{\mathrm{val}}
=
\left\{
\exists\,\ell\in\{1,\ldots,L\}:
\widehat{\mathbf{Y}}_{\ell}
\neq
\mathbf{Y}_{\ell}
\right\}
\cap
\left\{
\mathsf{Accept}=1
\right\}.
\label{eq:validation-failure-event}
\end{equation}

\begin{proposition}[Soundness of the verification layer]
\label{prop:polar-validation-soundness}
Suppose that at most $f^{\mathrm{wrk}}$ workers return incorrect response
bundles. Suppose further that each such worker commits to its bundle before
obtaining information about the $\gamma$ secret verification vectors. If the
committee outputs only when the accepted response set belongs to
$\mathfrak{D}(\mathbf{G})$, then
\begin{equation}
\Pr\!\left[
\mathcal{E}_{\mathrm{val}}
\right]
\leq
f^{\mathrm{wrk}}LQ^{-\gamma}
+
\epsilon_{\mathrm{auth}},
\label{eq:polar-validation-soundness}
\end{equation}
where $\epsilon_{\mathrm{auth}}$ is the aggregate failure probability of the
authentication, commitment, secret-sharing, native-finality, and
committee-certification mechanisms used by the validation instance.
\end{proposition}

The proof is given in
Appendix~\ref{app:proof-validation-soundness}. Polar coding handles
unavailable or rejected responses, whereas the hidden linear checks bound the
probability that an incorrect response enters the decoder.

\begin{corollary}[Verification-check budget]
\label{cor:verification-budget}
Let $\epsilon_{\mathrm{tar}}>\epsilon_{\mathrm{auth}}$ be a target
per-instance false-acceptance probability. If
$f^{\mathrm{wrk}}L>0$ and
\begin{equation}
\gamma
\geq
\left\lceil
\log_Q
\left(
\frac{f^{\mathrm{wrk}}L}
{\epsilon_{\mathrm{tar}}-\epsilon_{\mathrm{auth}}}
\right)
\right\rceil,
\label{eq:verification-budget}
\end{equation}
then
$\Pr[\mathcal{E}_{\mathrm{val}}]\leq\epsilon_{\mathrm{tar}}$.
If $f^{\mathrm{wrk}}L=0$, the same conclusion holds with
$\epsilon_{\mathrm{tar}}\geq\epsilon_{\mathrm{auth}}$.
\end{corollary}

We now characterize the confirmation threshold. For
$\mathcal{S}\subseteq\mathcal{N}$, let
\[
\omega(\mathcal{S})
=
\sum_{j\in\mathcal{S}}
\omega_j,
\]
and define
\begin{equation}
\mathfrak{Q}_{\eta}
=
\left\{
\mathcal{S}\subseteq\mathcal{N}:
\omega(\mathcal{S})\geq\eta
\right\}.
\label{eq:confirmation-quorums}
\end{equation}
Because membership and weights remain fixed within a checkpoint interval,
the minimum intersection weight is
\begin{equation}
\chi_{\boldsymbol{\omega}}(\eta)
=
\min_{\mathcal{S},\mathcal{S}'\in\mathfrak{Q}_{\eta}}
\omega
\left(
\mathcal{S}\cap\mathcal{S}'
\right).
\label{eq:minimum-quorum-intersection}
\end{equation}

\begin{proposition}[Exact weighted-quorum condition]
\label{prop:tight-quorum}
Suppose that the adversary may corrupt any chain set
$\mathcal{M}\subseteq\mathcal{N}$ satisfying
$\omega(\mathcal{M})\leq\rho$, and that an honest chain never supports two
conflicting blocks. Then conflicting blocks cannot both be confirmed for any
admissible corrupted set if and only if
\begin{equation}
\chi_{\boldsymbol{\omega}}(\eta)
>
\rho.
\label{eq:exact-quorum-safety}
\end{equation}
For a fixed honest-chain set $\mathcal{N}_{\mathrm{h}}$, confirmation without
Byzantine participation is possible if and only if
\begin{equation}
\omega
\left(
\mathcal{N}_{\mathrm{h}}
\right)
\geq
\eta.
\label{eq:exact-quorum-liveness}
\end{equation}
\end{proposition}

The necessity statement is with respect to the partially synchronous model:
before the global stabilization time, the adversary may delay information
between honest chains supporting different branches. The proof is given in
Appendix~\ref{app:proof-tight-quorum}.

\begin{corollary}[Weight-independent threshold region]
\label{cor:universal-threshold}
If
\begin{equation}
\frac{1+\rho}{2}
<
\eta
\leq
1-\rho,
\label{eq:universal-threshold-region}
\end{equation}
then conflicting-block safety and confirmation by honest chains hold for
every weight vector satisfying
$\omega(\mathcal{N}_{\mathrm{a}})\leq\rho$. The interval in
\eqref{eq:universal-threshold-region} is nonempty if and only if
$\rho<1/3$.
\end{corollary}

\begin{corollary}[Uniform-weight threshold specialization]
\label{cor:uniform-weight-threshold}
Suppose $\omega_j=1/N$ for all $j$ and at most $f$ chains are Byzantine.
Let
\[
q_{\eta}
=
\lceil \eta N\rceil
\]
be the minimum number of distinct issuers required for confirmation. Then
conflicting-block safety holds if and only if
\begin{equation}
2q_{\eta}-N>f,
\label{eq:uniform-safety-threshold}
\end{equation}
and confirmation by honest chains alone is possible if and only if
\begin{equation}
q_{\eta}\leq N-f.
\label{eq:uniform-liveness-threshold}
\end{equation}
\end{corollary}

\subsection{Cross-Layer Stability, Confirmation, and Settlement}
\label{subsec:cross-layer-analysis}

Consider potential block proposals from honest chain $j$ arriving according
to a Poisson process of rate $\lambda_j$. The proposal process is restricted
to locally admissible outgoing batches for which an honest committee quorum
is available. Let $\tau_{\mathrm{e}}$ be the validation deadline.

For the configured parent budget $K$, define
\begin{equation}
\Psi_j\!\left(K,R_j^{\mathrm{code}}\right)
=
\Pr\!\left[
T_{\mathrm{rec},j}^{(K+1)}
\leq
\tau_{\mathrm{e}}
\right].
\label{eq:deadline-recovery-probability}
\end{equation}
By Proposition~\ref{prop:recovery-time-law},
\begin{align}
\Psi_j\!\left(K,R_j^{\mathrm{code}}\right)
&=
\sum_{\mathcal{A}\in
\mathfrak{D}(\mathbf{G}_j)}
\prod_{i\in\mathcal{A}}
F_{j,i}^{(K+1)}(\tau_{\mathrm{e}})
\nonumber\\
&\hspace{2.2cm}\times
\prod_{i\in[\bar n_j]\setminus\mathcal{A}}
\left(
1-
F_{j,i}^{(K+1)}(\tau_{\mathrm{e}})
\right).
\label{eq:deadline-recovery-expanded}
\end{align}

Let $\mathcal{E}_{\mathrm{val},j}$ be the validation-failure event for chain
$j$, and define the correct-completion event
\begin{equation}
\mathcal{C}_j\!\left(K,R_j^{\mathrm{code}}\right)
=
\left\{
T_{\mathrm{rec},j}^{(K+1)}
\leq
\tau_{\mathrm{e}}
\right\}
\cap
\mathcal{E}_{\mathrm{val},j}^{\mathrm{c}}.
\label{eq:correct-completion-event}
\end{equation}
Its probability is
\begin{equation}
\theta_j\!\left(K,R_j^{\mathrm{code}}\right)
=
\Pr\!\left[
\mathcal{C}_j\!\left(K,R_j^{\mathrm{code}}\right)
\right].
\label{eq:correct-completion-probability}
\end{equation}
No independence between recovery and verification failure is required.
Proposition~\ref{prop:polar-validation-soundness} gives
\begin{equation}
\theta_j\!\left(K,R_j^{\mathrm{code}}\right)
\geq
\left[
\Psi_j\!\left(K,R_j^{\mathrm{code}}\right)
-
\epsilon_{\mathrm{ver},j}
\right]_{+},
\label{eq:completion-probability-lower-bound}
\end{equation}
where
\[
\epsilon_{\mathrm{ver},j}
=
f_j^{\mathrm{wrk}}(K+1)Q^{-\gamma_j}
+
\epsilon_{\mathrm{auth},j}
\]
and $[x]_{+}=\max\{x,0\}$.

Assume that correct-completion indicators are independent across potential
proposals and independent of proposal arrival times. Poisson thinning then
gives the effective honest issuance rate
\begin{equation}
\nu_{\mathrm{h}}\!\left(K,\mathbf{R}^{\mathrm{code}}\right)
=
\sum_{j\in\mathcal{N}_{\mathrm{h}}}
\lambda_j
\theta_j\!\left(K,R_j^{\mathrm{code}}\right).
\label{eq:effective-honest-rate}
\end{equation}
Adversarial blocks arrive according to an independent Poisson process of rate
$\lambda_{\mathrm{a}}$.

Let $L_t$ denote the number of public tips at the beginning of update
interval $t$. Rates are normalized per update interval. Every successfully
issued honest block selects
\[
K_{\ell}
=
\min\{K,\ell\}
\]
distinct public tips when $L_t=\ell$. The selected tips are conditionally
uniform without replacement, and selections made by different honest blocks
are conditionally independent. Blocks issued during interval $t$ become
eligible for selection only in interval $t+1$.

Let $\widehat{H}_t$ and $A_t$ denote the numbers of correctly issued honest
and adversarial blocks during interval $t$, respectively. Let $D_t$ be the
number of distinct existing public tips approved by the honest blocks.
Adversarial blocks are conservatively modeled as approving no public tip.
The public-tip process satisfies
\begin{equation}
L_{t+1}
=
L_t-D_t+\widehat{H}_t+A_t.
\label{eq:tip-process}
\end{equation}
For $\ell\geq1$, its conditional drift is
\begin{equation}
\delta(\ell)
=
\nu_{\mathrm{h}}
+
\lambda_{\mathrm{a}}
-
\ell
\left[
1-
\exp\!\left(
-\frac{\nu_{\mathrm{h}}K_{\ell}}{\ell}
\right)
\right].
\label{eq:exact-tip-drift}
\end{equation}

\begin{theorem}[Coded-validation-induced DAG stability]
\label{thm:cross-layer-stability}
Assume that $K\geq2$,
$\nu_{\mathrm{h}}(K,\mathbf{R}^{\mathrm{code}})>0$, and that the Markov
chain defined by~\eqref{eq:tip-process} is irreducible and aperiodic.

If
\begin{equation}
\lambda_{\mathrm{a}}
<
(K-1)
\nu_{\mathrm{h}}\!\left(K,\mathbf{R}^{\mathrm{code}}\right),
\label{eq:dag-stability-condition}
\end{equation}
then $\{L_t\}_{t\geq0}$ is positive recurrent and has a unique stationary
distribution with finite first moment.

If
\begin{equation}
\lambda_{\mathrm{a}}
>
(K-1)
\nu_{\mathrm{h}}\!\left(K,\mathbf{R}^{\mathrm{code}}\right),
\label{eq:dag-instability-condition}
\end{equation}
then
\begin{equation}
\liminf_{t\rightarrow\infty}
\frac{L_t}{t}
\geq
\lambda_{\mathrm{a}}
-
(K-1)
\nu_{\mathrm{h}}\!\left(K,\mathbf{R}^{\mathrm{code}}\right)
>
0
\quad
\text{almost surely}.
\label{eq:linear-tip-divergence}
\end{equation}

In particular, the computable condition
\begin{equation}
\lambda_{\mathrm{a}}
<
(K-1)
\sum_{j\in\mathcal{N}_{\mathrm{h}}}
\lambda_j
\left[
\Psi_j\!\left(K,R_j^{\mathrm{code}}\right)
-
\epsilon_{\mathrm{ver},j}
\right]_{+}
\label{eq:explicit-cross-layer-stability}
\end{equation}
is sufficient for positive recurrence.
No recurrence claim is made at the equality boundary.
\end{theorem}

The proof is given in Appendix~\ref{app:proof-dag-stability}. The explicit
condition~\eqref{eq:explicit-cross-layer-stability} is the principal
cross-layer relation: the complete heterogeneous worker-response law enters
the DAG stability condition through
$\Psi_j(K,R_j^{\mathrm{code}})$.

Let
\[
\lambda_{\mathrm{h}}
=
\sum_{j\in\mathcal{N}_{\mathrm{h}}}
\lambda_j,
\qquad
\bar{\theta}\!\left(K,\mathbf{R}^{\mathrm{code}}\right)
=
\frac{
\nu_{\mathrm{h}}\!\left(K,\mathbf{R}^{\mathrm{code}}\right)
}{
\lambda_{\mathrm{h}}
},
\]
where $\lambda_{\mathrm{h}}>0$. Define the potential-issuance adversarial
fraction
\begin{equation}
\mu
=
\frac{
\lambda_{\mathrm{a}}
}{
\lambda_{\mathrm{a}}+\lambda_{\mathrm{h}}
}.
\label{eq:potential-spam-fraction}
\end{equation}
Condition~\eqref{eq:dag-stability-condition} is equivalent to
\begin{equation}
\mu
<
\mu_{\mathrm{crit}}
\left(
K,\bar{\theta}
\right)
\triangleq
\frac{
(K-1)\bar{\theta}(K,\mathbf{R}^{\mathrm{code}})
}{
1+
(K-1)\bar{\theta}(K,\mathbf{R}^{\mathrm{code}})
}.
\label{eq:cross-layer-critical-spam}
\end{equation}
Thus, imperfect coded-validation completion contracts the stable adversarial
issuance region. When $\bar{\theta}=1$,
\eqref{eq:cross-layer-critical-spam} reduces to
$\mu_{\mathrm{crit}}=(K-1)/K$.

Consider a valid block $B$ issued after global stabilization in the stable
regime. For each honest chain $j$, let
$p_j^{\mathrm{sup}}(B)\in(0,1]$ be the probability that, in an interval in
which $j$ has not yet supported $B$, it issues a valid block whose past cone
contains $B$. Let $\tau_j(B)$ be the first such interval. Assume that
$\{\tau_j(B)\}_{j\in\mathcal{N}_{\mathrm{h}}}$
are independent geometric random variables with fixed parameters
$\{p_j^{\mathrm{sup}}(B)\}$.

Define
\begin{equation}
T_{\mathrm{dag}}(B)
=
\inf
\left\{
r\geq1:
\sum_{j\in\mathcal{N}_{\mathrm{h}}}
\omega_j
\mathds{1}
\left\{
\tau_j(B)\leq r
\right\}
\geq
\eta
\right\}.
\label{eq:weighted-confirmation-time}
\end{equation}

\begin{proposition}[Weighted confirmation-time law]
\label{prop:weighted-confirmation-time}
For every integer $r\geq0$,
\begin{equation}
\Pr\!\left[
T_{\mathrm{dag}}(B)>r
\right]
=
\sum_{\substack{
\mathcal{S}\subseteq\mathcal{N}_{\mathrm{h}}\\
\omega(\mathcal{S})<\eta
}}
\prod_{j\in\mathcal{S}}
\left[
1-
\left(
1-p_j^{\mathrm{sup}}(B)
\right)^r
\right]
\prod_{j\in\mathcal{N}_{\mathrm{h}}\setminus\mathcal{S}}
\left(
1-p_j^{\mathrm{sup}}(B)
\right)^r.
\label{eq:exact-confirmation-tail}
\end{equation}
Moreover,
\begin{equation}
\mathbb{E}
\left[
T_{\mathrm{dag}}(B)
\right]
=
\sum_{r=0}^{\infty}
\Pr\!\left[
T_{\mathrm{dag}}(B)>r
\right].
\label{eq:exact-confirmation-mean}
\end{equation}
The expectation is finite if an honest confirmation quorum
$\mathcal{H}$ exists such that
$p_j^{\mathrm{sup}}(B)>0$ for every $j\in\mathcal{H}$.
\end{proposition}

The proof is given in Appendix~\ref{app:proof-confirmation-time}.

Finally, consider an execution prefix containing $H$ validation instances
performed by honest committees. Let $\mathcal{F}_H$ be the ordered set of
blocks confirmed during the prefix, and let $\mathbf{b}^{(r)}$ be the state
after applying the first $r$ blocks in the deterministic checkpoint order.
For transaction identifier $\iota$, let $N_H(\iota)$ be its number of
destination applications.

Define
\begin{align}
\mathcal{E}_{\mathrm{neg}}^{(H)}
&=
\left\{
\exists\,r\leq|\mathcal{F}_H|:
\mathbf{b}^{(r)}
\nsucceq
\mathbf{0}
\right\},
\label{eq:negative-balance-event}\\
\mathcal{E}_{\mathrm{con}}^{(H)}
&=
\left\{
\exists\,r\leq|\mathcal{F}_H|:
\mathbf{1}^{\mathsf T}\mathbf{b}^{(r)}
\neq
\mathbf{1}^{\mathsf T}\mathbf{b}^{(0)}
\right\},
\label{eq:conservation-event}\\
\mathcal{E}_{\mathrm{conf}}^{(H)}
&=
\left\{
\exists\,B,B'\in\mathcal{F}_H:
B\perp B'
\right\},
\label{eq:conflicting-confirmation-event}\\
\mathcal{E}_{\mathrm{dup}}^{(H)}
&=
\left\{
\exists\,\iota:
N_H(\iota)>1
\right\},
\label{eq:duplicate-application-event}\\
\mathcal{E}_{\mathrm{code}}^{(H)}
&=
\left\{
\begin{aligned}
\exists\,(j,t,i)\text{ such that }\;&
j\in\mathcal{N}_{\mathrm{h}},\
i\in\mathcal{W}_j(t),\
i\text{ is honest},\\
&
\widetilde{\mathbf{b}}_{j,i}(t)
\neq
\sum_{\ell=1}^{k_j(t)}
\mathbf{G}_j(t)[i,\ell]
\mathbf{b}_{j,\ell}(t)
\end{aligned}
\right\}.
\label{eq:coded-state-failure-event}
\end{align}
Let
\begin{equation}
\mathcal{E}_{\mathrm{set}}^{(H)}
=
\mathcal{E}_{\mathrm{neg}}^{(H)}
\cup
\mathcal{E}_{\mathrm{con}}^{(H)}
\cup
\mathcal{E}_{\mathrm{conf}}^{(H)}
\cup
\mathcal{E}_{\mathrm{dup}}^{(H)}
\cup
\mathcal{E}_{\mathrm{code}}^{(H)}.
\label{eq:settlement-failure-event}
\end{equation}

\begin{theorem}[Cross-layer settlement guarantee]
\label{thm:cross-layer-settlement}
Suppose that the adapter interface and one-outstanding-sequence restriction
of Section~\ref{sec:system-model} hold. Suppose further that
\eqref{eq:exact-quorum-safety} holds and that, before contributing support,
every honest chain independently verifies the deterministic admissibility of
the supported block against its referenced checkpoint.

For validation instance $h$, let
$\epsilon_{\mathrm{ver}}^{(h)}$
be the bound in~\eqref{eq:polar-validation-soundness}. Then
\begin{equation}
\Pr\!\left[
\mathcal{E}_{\mathrm{set}}^{(H)}
\right]
\leq
\sum_{h=1}^{H}
\epsilon_{\mathrm{ver}}^{(h)}.
\label{eq:end-to-end-safety-bound}
\end{equation}

If, in addition,
\eqref{eq:exact-quorum-liveness} and
\eqref{eq:dag-stability-condition} hold, and an honest confirmation quorum
exists whose members have strictly positive support probabilities for $B$,
then every valid block $B$ issued by an honest chain is confirmed almost
surely. Its expected lock-to-release latency is finite whenever the
validation and destination-application stages have finite means.
\end{theorem}

The proof is given in
Appendix~\ref{app:proof-cross-layer-settlement}. The theorem applies to the
escrow-backed transfer and deterministic linear-state model defined in
Section~\ref{sec:system-model}; it does not assert correctness for arbitrary
smart-contract execution.

Let $T_{\mathrm{val}}(B)$ be the delay from native finalization of the source
lock event to block issuance, and let $T_{\mathrm{apply}}(B)$ be the delay
from confirmation to completion of all destination applications. Then
\begin{equation}
T_{\mathrm{fin}}(B)
=
T_{\mathrm{val}}(B)
+
T_{\mathrm{dag}}(B)
+
T_{\mathrm{apply}}(B).
\label{eq:end-to-end-latency}
\end{equation}
The validation delay decomposes as
\begin{equation}
T_{\mathrm{val}}(B)
=
T_{\mathrm{ctl}}(B)
+
T_{\mathrm{rec}}^{(L)}(B)
+
T_{\mathrm{cert}}(B),
\label{eq:validation-latency-decomposition}
\end{equation}
where $T_{\mathrm{ctl}}$ contains query construction and task dissemination,
and $T_{\mathrm{cert}}$ contains committee certification and block
publication.

Therefore,
\begin{equation}
\mathbb{E}
\left[
T_{\mathrm{fin}}(B)
\right]
=
\mathbb{E}
\left[
T_{\mathrm{val}}(B)
\right]
+
\mathbb{E}
\left[
T_{\mathrm{dag}}(B)
\right]
+
\mathbb{E}
\left[
T_{\mathrm{apply}}(B)
\right].
\label{eq:end-to-end-mean}
\end{equation}
If these three components are mutually independent, then
\begin{equation}
F_{\mathrm{fin}}
=
F_{\mathrm{val}}
*
F_{\mathrm{dag}}
*
F_{\mathrm{apply}},
\label{eq:end-to-end-convolution}
\end{equation}
where $*$ denotes convolution.

\section{Simulation Setup and Performance Evaluation}
\label{sec:evaluation}

We evaluate SPID using a prototype-assisted discrete-event simulator that
implements the adapter state machine, persistent coded-state maintenance,
worker-response verification, deterministic DAG parent selection, and
distinct-chain weighted confirmation described in
Sections~\ref{sec:spid-protocol}--\ref{sec:security}. The evaluation addresses
five questions: how coded validation behaves under heterogeneous worker
delays; how event-certified execution compares with periodic polling; how
validation and confirmation throughput scale; whether the observed public-tip
process follows the predicted stability boundary; and how the coding rate
and parent budget $K$ interact.

\subsection{Experimental Methodology}
\label{subsec:evaluation-setup}

The participating chains are instantiated using the Substrate framework
provided by the Polkadot SDK~\cite{PolkadotSDKDocs} and executed in Docker
containers~\cite{DockerDocs}. GoShimmer~\cite{GoShimmerRepo} provides the
peer-to-peer dissemination and DAG-networking layer used in the prototype.
Because GoShimmer is an archived research prototype rather than a maintained
production platform, SPID-Chain does not rely on its native confirmation
mechanism.

Each chain maintains its native ledger, source reservations, gateway sequence
numbers, replay-protection state, committee transcript, and local DAG view.
Each physical worker stores one persistent coded-state fragment and processes
a response bundle containing one coded result for every item in the joint
validation workload. Following checkpoint confirmation, a worker receives
only its coded component of the confirmed state increment. Missing responses
and responses rejected by the verification layer are treated as erasures.

The default deployment contains $N=10$ chains with equal attestation weights.
Two chains are Byzantine, so $\rho=0.2$, and the confirmation threshold is
$\eta=0.67$. Consequently, confirmation requires support from at least seven
distinct chains. The resulting minimum intersection of two confirmation
quorums is $0.4>\rho$, while the aggregate honest weight is $0.8>\eta$.
Equivalently, Corollary~\ref{cor:uniform-weight-threshold} gives
$2\cdot7-10=4>2$ for safety and $7\leq8$ for honest-only liveness.

Each chain has a committee of
$n_j^{\mathrm{com}}=10$ members, with certificate threshold $q_j=7$ and
assumed bound $f_j^{\mathrm{com}}=3$. Unless varied explicitly, the system
uses $n=100$ physical workers, $k=50$ uncoded state blocks, parent budget
$K=2$, a validation deadline $\tau_{\mathrm{e}}=2$~s, a one-way propagation
delay of $100$~ms, and an available bandwidth of $20$~Mbit/s.

For SPID-Polar, the physical-worker count $n=100$ induces code length
$\bar n=128$. The first $100$ reliability-assigned codeword positions are
mapped to physical workers, while the remaining $28$ positions are permanent
virtual erasures. Thus, SPID stores $100$ coded fragments for $50$ uncoded
state blocks. The MDS baseline uses an $(100,50)$ Reed--Solomon code and also
stores $100$ coded fragments. Rep-2 stores two replicas of each of the
$50$ uncoded blocks. The uncoded baseline stores one copy of each block and
therefore assigns only $50$ workers per instance; the remaining workers are
idle. The comparison is consequently matched in logical workload and
physical worker pool, while the storage factors are $1$ for uncoded execution
and $2$ for Rep-2, MDS, and SPID-Polar. SPID and MDS are storage-matched over
physical workers but do not have the same nominal code length.

Worker heterogeneity consists of a persistent speed factor and an
instance-specific fluctuation. For worker $i$ and validation instance $r$,
the computation time is
\begin{equation}
T_{i,r}^{\mathrm{cmp}}
=
0.18\,S_i Z_{i,r}\,\si{\second}.
\label{eq:evaluation-worker-time}
\end{equation}
where
\[
S_i
\sim
\operatorname{Lognormal}
\left(
-\frac{\sigma_s^2}{2},
\sigma_s^2
\right),
\qquad
Z_{i,r}
\sim
\operatorname{Lognormal}
\left(
-\frac{\sigma_z^2}{2},
\sigma_z^2
\right),
\]
with $\sigma_s=0.22$ and $\sigma_z=0.18$. The value of $S_i$ remains fixed
throughout one independent run, whereas $Z_{i,r}$ is resampled for every
validation instance.

The communication delay of a worker response follows a nonnegative truncated
normal distribution with mean $100$~ms and standard deviation $12$~ms. A
worker is independently designated as a straggler with probability
$p_{\mathrm{str}}$. A straggling response receives an additional
exponentially distributed delay with mean $650$~ms, and $2.5\%$ of stragglers
do not return before the validation deadline.

Table~\ref{tab:evaluation-parameters} summarizes the default configuration.
Ordinary throughput experiments use a $12$-min measurement interval after
warm-up and are repeated for ten independent seeds. Long-run DAG experiments
use a $60$-min measurement interval. The coded-validation experiment contains
$50{,}000$ independent instances for every method and every value of
$p_{\mathrm{str}}$.

\begin{table}[t]
\centering
\caption{Default evaluation parameters.}
\label{tab:evaluation-parameters}
\small
\setlength{\tabcolsep}{6pt}
\begin{tabular}{@{}ll@{}}
\toprule
Parameter & Default value \\
\midrule
Participating chains, $N$ & $10$ \\
Byzantine chains & $2$ \\
Attestation weights & Uniform \\
Adversarial weight bound, $\rho$ & $0.20$ \\
Confirmation threshold, $\eta$ & $0.67$ \\
Committee size, $n_j^{\mathrm{com}}$ & $10$ \\
Committee threshold, $q_j$ & $7$ \\
Committee fault bound, $f_j^{\mathrm{com}}$ & $3$ \\
Physical workers per chain, $n$ & $100$ \\
Polar code length, $\bar n$ & $128$ \\
Uncoded state blocks, $k$ & $50$ \\
Default parent budget, $K$ & $2$ \\
Default straggler probability, $p_{\mathrm{str}}$ & $0.10$ \\
Validation deadline, $\tau_{\mathrm{e}}$ & $2$ s \\
One-way propagation delay & $100$ ms \\
Available bandwidth & $20$ Mbit/s \\
Standard block payload & $1$ MB \\
Heavy block payload & $2$ MB \\
Ordinary measurement interval & $12$ min \\
Long-run measurement interval & $60$ min \\
Independent throughput runs & $10$ \\
Validation instances per setting & $50{,}000$ \\
\bottomrule
\end{tabular}
\end{table}

Four validation schemes are compared. The uncoded scheme completes only after
all $k$ assigned shards return. Rep-2 assigns two workers to each uncoded
state block and accepts the first verified response from each pair. The MDS
scheme reconstructs from any $k$ accepted symbols of an $(n,k)$
Reed--Solomon code. SPID-Polar uses the reliability-aware information set in
Section~\ref{subsec:persistent-polar-state} and the rank condition in
\eqref{eq:polar-decodability}. All four schemes process the same logical
candidate workload and use the same worker-time model, network configuration,
deadline, and verification procedure.

\subsection{Results and Discussion}
\label{subsec:evaluation-results}

Table~\ref{tab:coded-validation-results} reports the mean recovery latency,
the $95$th-percentile latency, and the empirical deadline-completion
frequency. The latency includes query preparation, worker computation,
communication, verification, and decoding.

\begin{table}[t]
\centering
\caption{Coded-validation performance under worker straggling. A reported
completion frequency of $100.00\%$ means that no deadline miss was observed
among $50{,}000$ trials; it does not assert a true completion probability of
one.}
\label{tab:coded-validation-results}
\small
\setlength{\tabcolsep}{4.5pt}
\begin{tabular}{@{}lcrrr@{}}
\toprule
Method & $p_{\mathrm{str}}$ & Mean (ms) & P95 (ms) & Completion (\%) \\
\midrule
Uncoded
& $0.0$ & $368.5$ & $434.6$ & $100.00$ \\
& $0.1$ & $1272.8$ & $1910.0$ & $62.49$ \\
& $0.3$ & $1633.5$ & $1971.2$ & $24.23$ \\
& $0.5$ & $1746.9$ & $1978.8$ & $9.19$ \\
\midrule
Rep-2
& $0.0$ & $353.2$ & $407.4$ & $100.00$ \\
& $0.1$ & $509.5$ & $1058.0$ & $99.52$ \\
& $0.3$ & $983.0$ & $1682.6$ & $95.83$ \\
& $0.5$ & $1281.6$ & $1850.8$ & $88.57$ \\
\midrule
MDS
& $0.0$ & $550.1$ & $588.6$ & $100.00$ \\
& $0.1$ & $556.8$ & $597.1$ & $100.00$ \\
& $0.3$ & $576.8$ & $622.1$ & $100.00$ \\
& $0.5$ & $625.7$ & $696.7$ & $100.00$ \\
\midrule
SPID-Polar
& $0.0$ & $516.4$ & $556.2$ & $100.00$ \\
& $0.1$ & $524.0$ & $566.0$ & $100.00$ \\
& $0.3$ & $548.1$ & $596.4$ & $100.00$ \\
& $0.5$ & $621.0$ & $719.7$ & $100.00$ \\
\bottomrule
\end{tabular}
\end{table}

Without stragglers, uncoded execution and replication avoid coding overhead
and consequently have the lowest mean latency. This advantage disappears as
the probability of delayed responses increases. At $p_{\mathrm{str}}=0.1$,
the uncoded completion frequency falls to $62.49\%$, whereas no deadline miss
is observed for either coded scheme. Rep-2 remains effective under light
straggling but develops a substantially longer tail because every replica
pair must contribute at least one accepted response.

The MDS baseline has the strongest cardinality-based recovery property,
because any $k$ accepted symbols suffice. SPID-Polar instead depends on the
identities of the responding positions through
$\mathfrak{D}(\mathbf{G})$. Nevertheless, its structured encoding and
decoding produce a lower mean latency than MDS for
$p_{\mathrm{str}}\leq0.3$. At $p_{\mathrm{str}}=0.5$, the mean latencies are
comparable, while MDS has the smaller $95$th percentile. This behavior is
consistent with Proposition~\ref{prop:recovery-time-law}: the Polar recovery
law depends on the complete response-set distribution rather than on response
cardinality alone.

\begin{figure}[t]
\centering
\includegraphics[width=0.95\linewidth]
{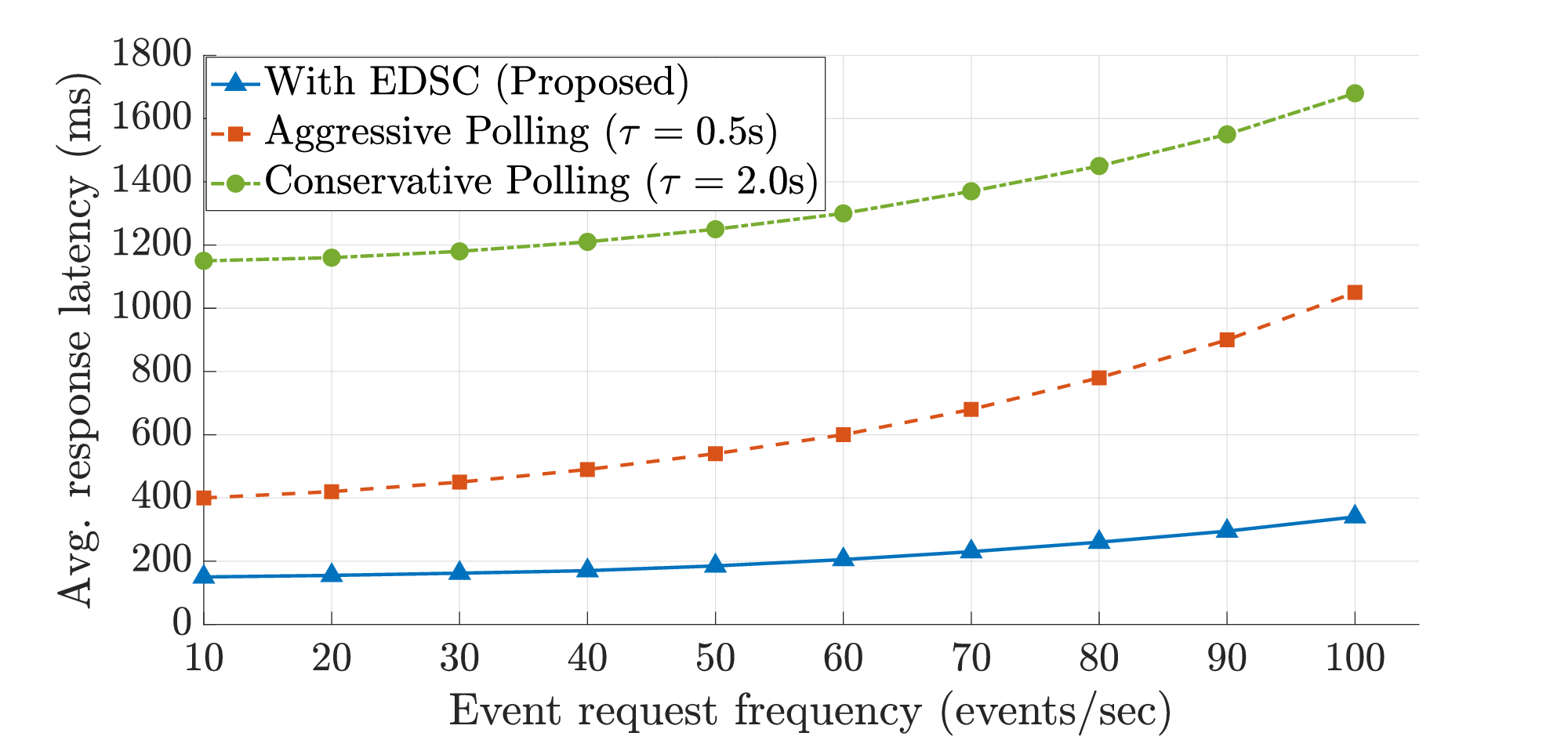}
\caption{Stage-response latency under event-certified execution and periodic
polling with intervals of $0.5$~s and $2$~s.}
\label{fig:ablation-edsc}
\end{figure}

Figure~\ref{fig:ablation-edsc} compares event-certified execution with
periodic polling at intervals of $0.5$~s and $2$~s. All configurations use
the same validation and state-transition logic and differ only in their
activation mechanism. Event certification avoids the residual delay incurred
while waiting for the next polling instant and also avoids repeated control
queries under high event rates.

Table~\ref{tab:throughput-scaling} reports validation and confirmed-block
throughput. Each entry is the sample mean and standard deviation over ten
independent runs. The worker-count sweep uses an offered load of
$14$ batches/min, while the chain-count sweep uses an offered load of
$12$ blocks/min.

\begin{table*}[t]
\centering
\caption{Validation and confirmed-block throughput. Unless varied explicitly,
$N=10$, $n=100$, $p_{\mathrm{str}}=0.1$, $K=2$, and $\eta=0.67$. Entries are
mean $\pm$ standard deviation in blocks/min.}
\label{tab:throughput-scaling}
\small
\setlength{\tabcolsep}{4pt}
\resizebox{\textwidth}{!}{%
\begin{tabular}{@{}llcccccccc@{}}
\toprule
Experiment & Method &
\multicolumn{8}{c}{Varied parameter and measured throughput} \\
\midrule

\multicolumn{10}{@{}l}{
\textit{Validation throughput versus offered load (batches/min)}}\\
& & $2$ & $4$ & $6$ & $8$ & $10$ & $12$ & $14$ & $16$ \\
SPID-Polar & &
$1.98\pm0.03$ &
$3.99\pm0.05$ &
$5.85\pm0.09$ &
$7.79\pm0.09$ &
$9.58\pm0.17$ &
$10.88\pm0.26$ &
$11.17\pm0.19$ &
$11.35\pm0.15$ \\
Uncoded & &
$1.95\pm0.04$ &
$3.83\pm0.06$ &
$5.63\pm0.12$ &
$6.80\pm0.15$ &
$7.49\pm0.10$ &
$7.79\pm0.11$ &
$7.90\pm0.19$ &
$7.93\pm0.12$ \\
\midrule

\multicolumn{10}{@{}l}{
\textit{Validation throughput versus physical workers, $n$}}\\
& & $20$ & $40$ & $60$ & $80$ & $100$ & $120$ & $160$ & -- \\
SPID-Polar & &
$5.16\pm0.10$ &
$7.37\pm0.14$ &
$8.88\pm0.22$ &
$10.19\pm0.30$ &
$11.16\pm0.26$ &
$11.43\pm0.35$ &
$11.27\pm0.24$ &
-- \\
Uncoded & &
$4.58\pm0.07$ &
$6.19\pm0.13$ &
$6.91\pm0.14$ &
$7.47\pm0.15$ &
$7.86\pm0.10$ &
$7.89\pm0.12$ &
$7.78\pm0.14$ &
-- \\
\midrule

\multicolumn{10}{@{}l}{
\textit{Confirmed-block throughput versus offered load (blocks/min)}}\\
& & $2$ & $4$ & $6$ & $8$ & $10$ & $12$ & $14$ & $16$ \\
SPID-Polar & &
$1.85\pm0.04$ &
$3.65\pm0.08$ &
$5.46\pm0.10$ &
$7.04\pm0.24$ &
$8.40\pm0.22$ &
$9.12\pm0.19$ &
$9.26\pm0.17$ &
$9.35\pm0.27$ \\
Uncoded & &
$1.75\pm0.05$ &
$3.40\pm0.08$ &
$4.84\pm0.12$ &
$5.67\pm0.18$ &
$6.22\pm0.17$ &
$6.26\pm0.11$ &
$6.39\pm0.12$ &
$6.28\pm0.17$ \\
\midrule

\multicolumn{10}{@{}l}{
\textit{Confirmed-block throughput versus participating chains, $N$}}\\
& & $4$ & $6$ & $8$ & $10$ & $12$ & $16$ & $20$ & -- \\
SPID-Polar & &
$4.04\pm0.15$ &
$6.05\pm0.18$ &
$7.72\pm0.30$ &
$9.16\pm0.22$ &
$10.16\pm0.20$ &
$10.50\pm0.29$ &
$10.51\pm0.27$ &
-- \\
Uncoded & &
$3.40\pm0.11$ &
$4.86\pm0.13$ &
$5.69\pm0.20$ &
$6.22\pm0.14$ &
$6.33\pm0.11$ &
$6.62\pm0.14$ &
$6.48\pm0.13$ &
-- \\
\bottomrule
\end{tabular}}
\end{table*}

At low offered loads, both methods process nearly every submitted batch.
SPID-Polar separates from uncoded execution once the offered load exceeds
approximately $6$ batches/min and saturates near $11.35$ validated
batches/min. The uncoded configuration saturates below $8$ batches/min
because every instance remains dependent on its slowest assigned worker.

Increasing the worker count improves SPID throughput up to approximately
$120$ workers. Beyond this point, the additional parallelism no longer
compensates for task dissemination, verification, and response aggregation.
The uncoded configuration saturates earlier because it cannot complete from
a verified subset of responses.

The validation advantage propagates to the confirmation layer. At an offered
load of $16$ blocks/min, SPID confirms approximately $9.35$ blocks/min,
compared with $6.28$ blocks/min under uncoded validation. Increasing the
number of chains initially accelerates accumulation of distinct-chain
support. The gain becomes small beyond approximately $16$ chains because
network dissemination and committee processing become the dominant
bottlenecks.

\begin{figure*}[t]
\centering
\includegraphics[width=0.98\textwidth]{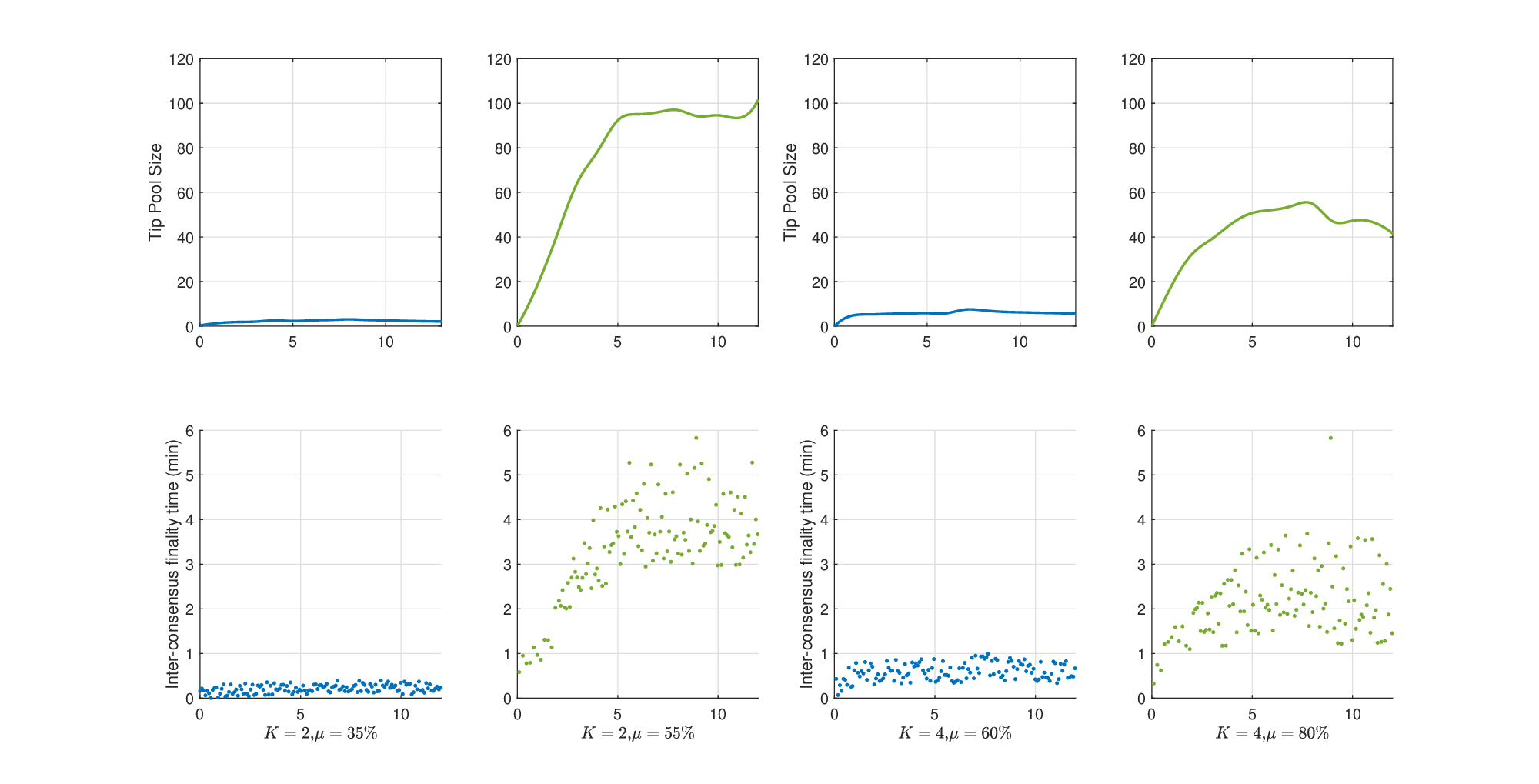}
\caption{Public-tip count and cross-chain finality below and above the
cross-layer stability boundary. The left pair uses $K=2$ and
$\mu\in\{0.35,0.55\}$; the right pair uses $K=4$ and
$\mu\in\{0.60,0.80\}$.}
\label{fig:dag-finality}
\end{figure*}

Figure~\ref{fig:dag-finality} evaluates the cross-layer stability relation.
For $K=2$, the adversarial issuance fractions are $\mu=0.35$ and
$\mu=0.55$; for $K=4$, they are $\mu=0.60$ and $\mu=0.80$. The observed
validation-completion probability is close to one in these experiments.
Equation~\eqref{eq:cross-layer-critical-spam} therefore predicts critical
fractions close to $0.5$ and $0.75$ for $K=2$ and $K=4$, respectively.

Below the corresponding boundary, the public-tip count remains bounded and
the finality samples remain concentrated. Above the boundary, the unapproved
tip population grows persistently and the finality distribution develops a
longer tail. The observed transition is consistent with
Theorem~\ref{thm:cross-layer-stability}.

\begin{figure}[t]
\centering
\includegraphics[width=0.95\linewidth]
{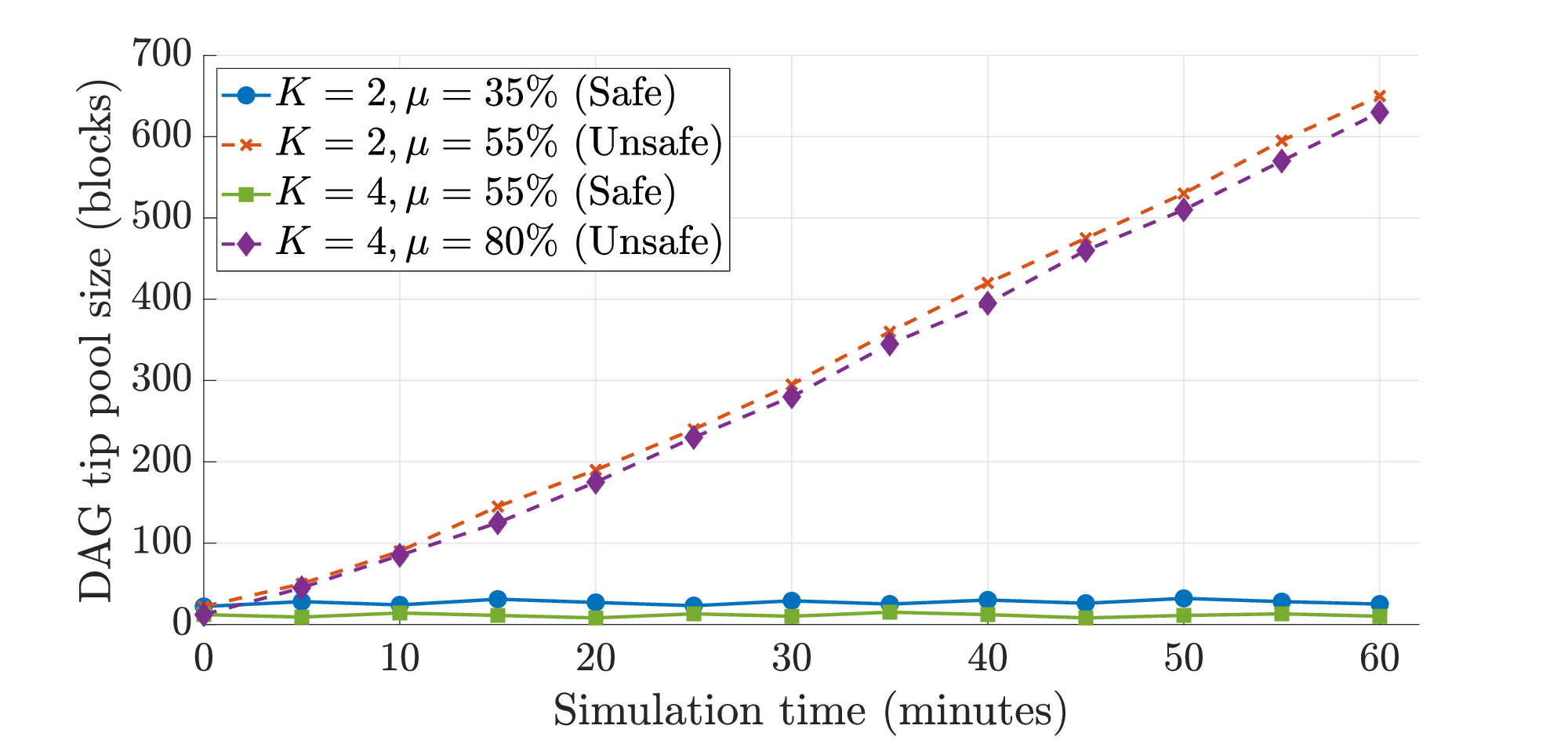}
\caption{Public-tip count over $60$ min for $K\in\{2,4\}$ and adversarial
issuance fractions below and above their predicted stability boundaries.}
\label{fig:longrun-stability}
\end{figure}

The $60$-min experiments in Fig.~\ref{fig:longrun-stability} distinguish
temporary tip-count fluctuations from persistent instability. For $K=2$ and
$\mu=0.55$, the public-tip population continues to increase because the
adversarial arrival rate exceeds the effective service capacity generated by
honest approvals. Increasing the parent budget to $K=4$ places the same
adversarial load inside the predicted stable region. At $\mu=0.80$, the
$K=4$ process again exhibits sustained growth.

\begin{figure}[t]
\centering
\includegraphics[width=0.95\linewidth]
{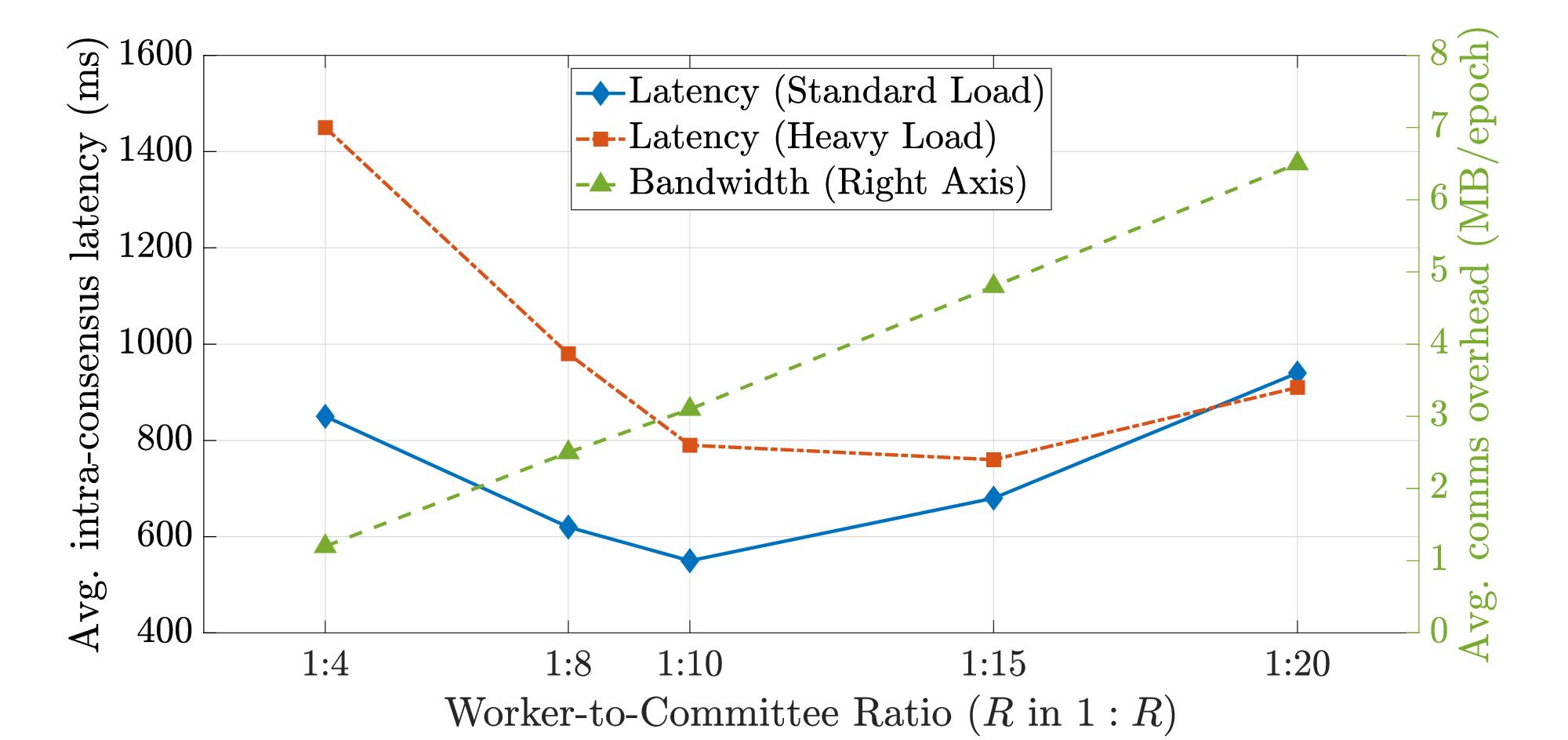}
\caption{Validation latency and communication volume versus the
committee-to-worker ratio for $1$-MB and $2$-MB workloads.}
\label{fig:ratio-optimization}
\end{figure}

Figure~\ref{fig:ratio-optimization} varies the committee-to-worker ratio. For
the $1$-MB workload, the minimum validation latency occurs near a
$1{:}10$ ratio. Smaller worker populations provide insufficient parallelism,
whereas larger populations increase dissemination and response-aggregation
costs. For the $2$-MB workload, the minimum shifts toward $1{:}15$ because
the larger task benefits from additional parallelism. The $1{:}10$
configuration remains near the minimum for both workloads and is used as the
default operating point.

\begin{figure}[t]
\centering
\includegraphics[width=0.95\linewidth]
{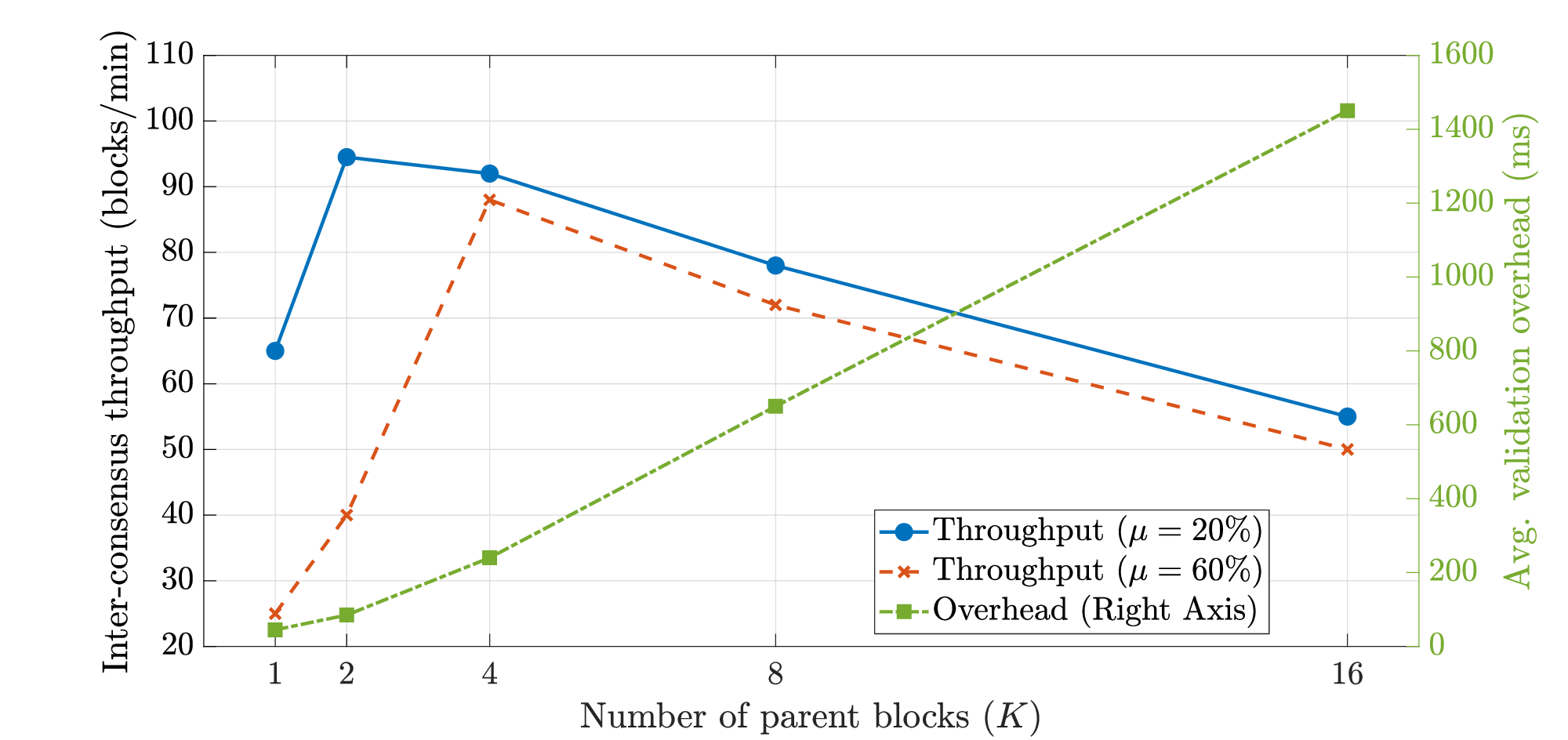}
\caption{Confirmed-block throughput and validation overhead versus $K$ under
nominal issuance, $\mu=0.20$, and adversarial issuance, $\mu=0.60$.}
\label{fig:sensitivity-k}
\end{figure}

Figure~\ref{fig:sensitivity-k} illustrates that $K$ cannot be selected
independently of the adversarial issuance rate. Under $\mu=0.20$, $K=2$
provides the largest confirmed throughput because validating additional
parents offers little stability benefit. Increasing $K$ enlarges the joint
workload from $K+1$ items and increases recovery and verification costs.

Under $\mu=0.60$, the $K=2$ configuration operates outside its predicted
stable region and confirmed throughput decreases. Setting $K=4$ moves the
system inside the stable region and restores bounded public-tip growth.
Larger values provide diminishing stability gains while continuing to
increase the coded workload. The preferred value of $K$ therefore depends
jointly on the worker completion law and the adversarial issuance rate, as
captured by~\eqref{eq:cross-layer-critical-spam}.

The results identify two operating regimes. When worker execution is nearly
homogeneous and adversarial issuance is low, uncoded execution and a small
parent budget avoid unnecessary redundancy. Under heterogeneous worker
delays, coded recovery prevents the slowest assigned workers from determining
the validation deadline. Under sustained adversarial issuance, increasing
$K$ enlarges the stable DAG region but also increases the coded-validation
workload. The coding configuration and parent budget must therefore be
selected jointly.

\section{Discussion and Limitations}
\label{sec:discussion}

\subsection{Deployment Implications}
\label{subsec:deployment-implications}

The analysis and evaluation indicate that the coded-validation configuration
and the directed acyclic graph (DAG) parameters cannot be selected
independently. Increasing the parent budget $K$ increases the number of
existing tips that an honest block can approve, thereby enlarging the
adversarial-load region in which the public-tip process is stable. The same
increase, however, enlarges the joint validation workload from one local
outgoing batch to at most $K+1$ state-transition items. This can reduce the
probability that an honest proposal completes validation before the deadline.
Similarly, reducing the code rate $R_j^{\mathrm{code}}$ increases redundancy
and tolerance to missing responses, but also increases storage,
communication, and verification costs. A deployment should therefore
estimate the empirical worker-response distributions, compute or approximate
the corresponding deadline-completion probabilities, and select $K$,
$R_j^{\mathrm{code}}$, and $\tau_{\mathrm e}$ jointly.

The confirmation threshold $\eta$ must also satisfy both safety and liveness
requirements. The exact condition
$\chi_{\boldsymbol{\omega}}(\eta)>\rho$ prevents two conflicting blocks from
obtaining confirmation quorums, whereas
$\omega(\mathcal{N}_{\mathrm h})\geq\eta$ permits confirmation without
Byzantine participation. Consequently, chain membership and attestation
weights should not be modified during an active checkpoint interval.
Reconfiguration should occur only through a confirmed checkpoint that fixes
the new membership, weights, committee parameters, code configuration, and
verification material before the next interval becomes active.

Persistent coding moves part of the validation cost from individual
transactions to checkpoint maintenance. Workers retain coded fragments of
the confirmed settlement state and receive only their coded components of
confirmed increments. This reduces repeated encoding of the complete state,
but it requires durable worker storage, authenticated checkpoint delivery,
and a re-encoding procedure whenever the worker population or generator
matrix changes. A production implementation must additionally define
recovery procedures for workers that miss several checkpoint updates.

\subsection{Scope and Trust Boundary}
\label{subsec:scope-limitations}

SPID-Chain is designed for deterministic, escrow-backed fungible transfers.
The state-transition model assumes that each accepted block induces a
deterministic debit vector and a deterministic credit vector whose total
values are equal. The resulting guarantees do not extend directly to
arbitrary smart-contract execution, non-deterministic external calls,
cross-chain contract composability, non-fungible assets, or application state
whose validity cannot be represented by the admissibility checks in
Section~\ref{sec:system-model}. Supporting such operations would require a
richer execution model and a corresponding extension of the coded workload
and settlement invariants.

The protocol also assumes at most one outstanding outgoing sequence per
issuer. This restriction prevents two independently validated blocks from the
same issuer from consuming the same checkpoint balance. Permitting multiple
concurrent outgoing sequences would require explicit balance partitioning,
reservation accounting, or a serializable concurrency-control mechanism.
The restriction is therefore a protocol limitation rather than a fundamental
property of coded validation.

Native-chain consensus remains outside the SPID-Chain trust boundary. The
protocol assumes that a native-finality certificate correctly establishes an
immutable source reservation and that the corresponding adapter exposes
consistent event, sequence, and replay information. SPID-Chain does not
correct failures of the native consensus protocol, compromised source-chain
governance, incorrect adapter code, stolen administrative keys, or fraudulent
asset issuance below the adapter interface. Similarly, exactly-once
destination execution depends on the destination adapter implementing the
check-record-apply operation atomically.

The committee guarantees require the threshold assumptions in
Section~\ref{sec:system-model}. If an adversary controls enough committee
members to produce an invalid threshold certificate, the certificate no
longer provides evidence of correct committee execution. The authentication
and certification failure term
$\epsilon_{\mathrm{auth}}$ explicitly accounts for failures of the external
cryptographic and committee mechanisms, but these mechanisms are not
constructed or proved within this paper.

\subsection{Verification Assumptions}
\label{subsec:verification-limitations}

The verification-soundness bound assumes that an incorrect worker response is
committed before the worker obtains information about the secret linear
verification vectors used for that instance. The vectors must therefore be
freshly generated, securely shared, or otherwise remain computationally
hidden until the response commitment becomes binding. Reusing the same secret
vectors across adaptively chosen responses while revealing acceptance
outcomes can leak information about the verification subspace and invalidate
the per-instance bound $Q^{-\gamma_j}$. A production implementation should
use fresh challenges for each validation instance or employ a
cryptographically secure homomorphic authentication mechanism with an
explicit adaptive-security proof.

The linear checks provide probabilistic integrity for the coded linear
workload; they do not provide privacy. Workers may observe their coded state
fragments, assigned update components, workload metadata, and timing
information. The present construction does not attempt to hide transaction
relationships, balances, participating chains, or access patterns.
Confidential settlement would require additional encryption,
secret-sharing, or zero-knowledge mechanisms and a separate analysis of their
interaction with coded recovery.

The field-size and canonical-embedding requirements are also substantive.
The prime field must be large enough to represent every admissible signed
state increment without modular ambiguity. Overflow, inconsistent integer
encoding, or implementation-dependent serialization could invalidate the
algebraic correspondence between the native settlement state and its coded
representation. These conditions must therefore be enforced by the adapter
and checkpoint formats.

\subsection{Analytical Limitations}
\label{subsec:analytical-limitations}

The exact recovery-time expression in
Proposition~\ref{prop:recovery-time-law} assumes mutually independent worker
completion times. Persistent hardware heterogeneity is represented through
worker-specific distributions, but common network congestion, shared
infrastructure failures, and correlated straggling can violate independence.
The algebraic decodability condition remains valid under correlation, but the
product-form distribution in
\eqref{eq:exact-recovery-cdf} must then be replaced by the joint distribution
of the response set or estimated empirically.

The DAG stability theorem is exact for the discrete-time model in
Section~\ref{sec:security}, which assumes Poisson issuance, interval-based
eligibility, conditionally uniform parent selection, and adversarial blocks
that do not remove public tips. Real deployments may exhibit bursty traffic,
correlated validation failures, nonuniform network views, and strategic
parent selection. The adversarial no-removal model is conservative with
respect to tip reduction, but other deviations can alter both the transient
tip distribution and the stability boundary. No recurrence classification is
claimed at
$\lambda_{\mathrm a}=(K-1)\nu_{\mathrm h}$.

The weighted confirmation-time distribution further assumes independent
geometric support times with fixed parameters. This is an analytical
specialization rather than a consequence of the protocol. In a deployed
network, support events can be correlated through common propagation delays,
shared committee members, and overlapping DAG views. The exact expression in
\eqref{eq:exact-confirmation-tail} should therefore be interpreted as a
tractable characterization under the stated model, while the quorum-safety
condition remains independent of this geometric approximation.

\subsection{Evaluation Limitations}
\label{subsec:evaluation-limitations}

The evaluation is based on a prototype-assisted discrete-event simulation
rather than a production deployment spanning independently administered
blockchains. Substrate-based chains, Docker containers, and the GoShimmer
networking prototype reproduce the principal execution, communication, and
DAG interactions needed by the study, but they do not capture the complete
operational environment of a public cross-chain service. In particular, the
evaluation does not model long-duration Internet partitions, economic
incentives, validator bribery, governance attacks, key compromise, denial of
service against gateways, or implementation vulnerabilities in production
bridge contracts.

The reported performance therefore characterizes the configured workload,
network model, worker-delay distributions, and adversarial issuance process.
Different state sizes, bandwidth distributions, cryptographic libraries,
committee implementations, and native-chain finality delays can change the
absolute throughput and latency. The evaluation is intended to test the
predicted cross-layer relationships and comparative behavior of the
validation schemes, not to establish a universal production-capacity bound.

\section{Conclusion}
\label{sec:conclusion}

This paper introduced SPID-Chain, a cross-chain settlement architecture that
combines adapter-based source reservations, persistent Polar-coded state
validation, Byzantine-response verification, and weighted directed acyclic
graph confirmation. The design preserves the autonomy of participating
blockchains and does not require modification of their native consensus
protocols. Its scope is restricted to deterministic, escrow-backed fungible
transfers and does not assume support for arbitrary cross-chain
smart-contract execution. The analysis established how worker heterogeneity
and coded-state recoverability determine the probability of completing a
validation instance before its deadline, how this probability controls the
effective honest block-issuance rate, and how the resulting issuance rate
determines the stability region of the settlement graph. Under the stated
committee, quorum, adapter, and liveness assumptions, SPID-Chain prevents
conflicting confirmations, negative settlement balances, asset creation or
destruction, duplicate destination execution, and divergence between the
confirmed state and the coded fragments maintained by honest workers. The
evaluation indicates that coded validation is most beneficial under
heterogeneous worker delays and response unavailability, while increasing the
DAG parent budget enlarges the stable adversarial-load region at the cost of
a larger validation workload. These results show that coded validation and
settlement-graph control must be configured jointly rather than optimized
independently. Future work will consider multiple concurrent outgoing
sequences, formally verified dynamic weight reconfiguration, verification
mechanisms secure under repeated adaptive interaction, and broader classes of
deterministic cross-chain state transitions.

\appendix

\section{Proofs of the Recovery-Time Results}
\label{app:proof-recovery-law}

\begin{proof}[Proof of Proposition~\ref{prop:recovery-time-law}]
Fix $\tau\geq0$. For each
$\mathcal{A}\subseteq[\bar n]$, define
\[
\mathcal{Z}_{\mathcal{A}}(\tau)
=
\left\{
\mathcal{A}^{(L)}(\tau)=\mathcal{A}
\right\}.
\]
The collection
$\{\mathcal{Z}_{\mathcal{A}}(\tau):
\mathcal{A}\subseteq[\bar n]\}$
partitions the sample space. Independence of the response times gives
\begin{equation}
\Pr\!\left[
\mathcal{Z}_{\mathcal{A}}(\tau)
\right]
=
\prod_{i\in\mathcal{A}}
F_i^{(L)}(\tau)
\prod_{i\in[\bar n]\setminus\mathcal{A}}
\left(
1-F_i^{(L)}(\tau)
\right).
\label{eq:proof-response-set-probability}
\end{equation}

The family $\mathfrak{D}(\mathbf{G})$ is upward closed. If
$\mathcal{A}\in\mathfrak{D}(\mathbf{G})$ and
$\mathcal{A}\subseteq\mathcal{A}'$, then
\[
k
=
\operatorname{rank}
\left(
\mathbf{G}_{\mathcal{A}}
\right)
\leq
\operatorname{rank}
\left(
\mathbf{G}_{\mathcal{A}'}
\right)
\leq
k.
\]
Hence,
$\mathcal{A}'\in\mathfrak{D}(\mathbf{G})$.

The response process is nondecreasing under set inclusion:
$\mathcal{A}^{(L)}(\tau_1)\subseteq
\mathcal{A}^{(L)}(\tau_2)$ whenever $\tau_1\leq\tau_2$.
Consequently,
\begin{equation}
\left\{
T_{\mathrm{rec}}^{(L)}
\leq
\tau
\right\}
=
\left\{
\mathcal{A}^{(L)}(\tau)
\in
\mathfrak{D}(\mathbf{G})
\right\}.
\label{eq:proof-recovery-event-equivalence}
\end{equation}
Indeed, membership of
$\mathcal{A}^{(L)}(\tau)$ in
$\mathfrak{D}(\mathbf{G})$ implies that the stopping condition has been met
by time $\tau$. Conversely, if the response set at time $\tau$ is not
decodable, no earlier subset can be decodable because
$\mathfrak{D}(\mathbf{G})$ is upward closed.

Combining
\eqref{eq:proof-response-set-probability} and
\eqref{eq:proof-recovery-event-equivalence} yields
\[
\Pr\!\left[
T_{\mathrm{rec}}^{(L)}
\leq
\tau
\right]
=
\sum_{\mathcal{A}\in\mathfrak{D}(\mathbf{G})}
\prod_{i\in\mathcal{A}}
F_i^{(L)}(\tau)
\prod_{i\in[\bar n]\setminus\mathcal{A}}
\left(
1-F_i^{(L)}(\tau)
\right),
\]
which proves~\eqref{eq:exact-recovery-cdf}.

Since $T_{\mathrm{rec}}^{(L)}$ is nonnegative, the tail-integral identity
gives
\[
\mathbb{E}
\left[
T_{\mathrm{rec}}^{(L)}
\right]
=
\int_{0}^{\infty}
\Pr\!\left[
T_{\mathrm{rec}}^{(L)}>\tau
\right]
\mathrm{d}\tau.
\]
Under the stated finite-mean assumption, this is
\eqref{eq:exact-recovery-mean}.
\end{proof}

\begin{proof}[Proof of Corollary~\ref{cor:recovery-spectrum}]
For every virtual position
$i\in[\bar n]\setminus[n]$,
$F_i^{(L)}(\tau)=0$. Therefore, every nonzero term in
\eqref{eq:exact-recovery-cdf} corresponds to a response set
$\mathcal{A}\subseteq[n]$.

For a physical response set with $|\mathcal{A}|=r$,
\[
\prod_{i\in\mathcal{A}}
F_i^{(L)}(\tau)
\prod_{i\in[n]\setminus\mathcal{A}}
\left(
1-F_i^{(L)}(\tau)
\right)
=
\left(
F^{(L)}(\tau)
\right)^r
\left(
1-F^{(L)}(\tau)
\right)^{n-r}.
\]
There are exactly $a_r(\mathbf{G})$ decodable physical response sets of
cardinality $r$. Grouping the terms in
\eqref{eq:exact-recovery-cdf} according to their cardinality gives
\[
\Pr\!\left[
T_{\mathrm{rec}}^{(L)}
\leq
\tau
\right]
=
\sum_{r=k}^{n}
a_r(\mathbf{G})
\left(
F^{(L)}(\tau)
\right)^r
\left(
1-F^{(L)}(\tau)
\right)^{n-r},
\]
which proves~\eqref{eq:spectrum-recovery-cdf}.
\end{proof}

\section{Proof of Verification Soundness}
\label{app:proof-validation-soundness}

\begin{proof}[Proof of Proposition~\ref{prop:polar-validation-soundness}]
Fix a Byzantine worker $i$ and workload item
$\ell\in\{1,\ldots,L\}$. Let
$\widetilde{\mathbf{y}}_{i,\ell}$ be the correct coded row and
$\widehat{\mathbf{y}}_{i,\ell}$ the committed row. If the committed row is
incorrect, define
\[
\mathbf{e}_{i,\ell}
=
\widehat{\mathbf{y}}_{i,\ell}
-
\widetilde{\mathbf{y}}_{i,\ell}
\neq
\mathbf{0}.
\]

For verification vector $\mathbf{r}_{i,h}$, the incorrect row satisfies the
corresponding check only if
\[
\left\langle
\mathbf{r}_{i,h},
\mathbf{e}_{i,\ell}
\right\rangle
=
0.
\]
Conditioned on the adversarial view at the commitment time,
$\mathbf{e}_{i,\ell}$ is fixed and nonzero. The mapping
\[
\varphi_{\mathbf{e}_{i,\ell}}:
\mathbb{F}_Q^{m}
\longrightarrow
\mathbb{F}_Q,
\qquad
\mathbf{r}
\longmapsto
\left\langle
\mathbf{r},
\mathbf{e}_{i,\ell}
\right\rangle
\]
is a nonzero linear functional. Its kernel has cardinality $Q^{m-1}$.
Therefore,
\[
\Pr\!\left[
\left\langle
\mathbf{r}_{i,h},
\mathbf{e}_{i,\ell}
\right\rangle
=
0
\right]
=
Q^{-1}.
\]

The $\gamma$ verification vectors are mutually independent, so
\begin{equation}
\Pr\!\left[
\widehat{\mathbf{y}}_{i,\ell}
\text{ passes all }\gamma\text{ checks}
\right]
=
Q^{-\gamma}.
\label{eq:proof-row-false-acceptance}
\end{equation}
No independence among the error vectors chosen by different Byzantine
workers is required.

At most $f^{\mathrm{wrk}}L$ worker--item pairs can contain incorrect rows.
A union bound applied to
\eqref{eq:proof-row-false-acceptance} gives
\begin{equation}
\Pr\!\left[
\exists\,(i,\ell):
\widehat{\mathbf{y}}_{i,\ell}
\neq
\widetilde{\mathbf{y}}_{i,\ell}
\text{ and the row is accepted}
\right]
\leq
f^{\mathrm{wrk}}LQ^{-\gamma}.
\label{eq:proof-verification-union-bound}
\end{equation}

Let $\mathcal{E}_{\mathrm{auth}}$ be the aggregate failure event for the
authentication, commitment, secret-sharing, native-finality, and
committee-certification mechanisms involved in the validation instance.
By definition,
$\Pr[\mathcal{E}_{\mathrm{auth}}]\leq
\epsilon_{\mathrm{auth}}$.

On the complement of $\mathcal{E}_{\mathrm{auth}}$ and the event in
\eqref{eq:proof-verification-union-bound}, every accepted coded row is
correct. Let $\mathcal{A}$ be the accepted response set. Since the committee
outputs only when
$\mathcal{A}\in\mathfrak{D}(\mathbf{G})$,
the matrix $\mathbf{G}_{\mathcal{A}}$ has full column rank. The accepted
coded rows therefore uniquely determine every uncoded post-debit matrix, and
\[
\widehat{\mathbf{Y}}_{\ell}
=
\mathbf{Y}_{\ell},
\qquad
\ell=1,\ldots,L.
\]

Thus,
\[
\mathcal{E}_{\mathrm{val}}
\subseteq
\mathcal{E}_{\mathrm{auth}}
\cup
\left\{
\exists\text{ an accepted incorrect row}
\right\}.
\]
Combining this inclusion with
\eqref{eq:proof-verification-union-bound} proves
\[
\Pr\!\left[
\mathcal{E}_{\mathrm{val}}
\right]
\leq
f^{\mathrm{wrk}}LQ^{-\gamma}
+
\epsilon_{\mathrm{auth}}.
\]
\end{proof}

\begin{proof}[Proof of Corollary~\ref{cor:verification-budget}]
If $f^{\mathrm{wrk}}L=0$, then
Proposition~\ref{prop:polar-validation-soundness} gives
$\Pr[\mathcal{E}_{\mathrm{val}}]\leq\epsilon_{\mathrm{auth}}$, which is at
most $\epsilon_{\mathrm{tar}}$ by assumption. Otherwise,
\eqref{eq:verification-budget} implies
\[
Q^{-\gamma}
\leq
\frac{\epsilon_{\mathrm{tar}}-\epsilon_{\mathrm{auth}}}{f^{\mathrm{wrk}}L}.
\]
Substituting this inequality into
\eqref{eq:polar-validation-soundness} gives
\[
\Pr[\mathcal{E}_{\mathrm{val}}]
\leq
f^{\mathrm{wrk}}LQ^{-\gamma}
+
\epsilon_{\mathrm{auth}}
\leq
\epsilon_{\mathrm{tar}}.
\]
\end{proof}

\section{Proofs of the Weighted-Quorum Results}
\label{app:proof-tight-quorum}

\begin{proof}[Proof of Proposition~\ref{prop:tight-quorum}]
We first establish sufficiency. Suppose
$\chi_{\boldsymbol{\omega}}(\eta)>\rho$ and, for contradiction, that
conflicting blocks $B$ and $B'$ are both confirmed. Let
$\mathcal{S},\mathcal{S}'\in\mathfrak{Q}_{\eta}$ be their respective
supporting-chain sets.

An honest chain does not support conflicting blocks. Therefore,
\[
\mathcal{S}\cap\mathcal{S}'
\subseteq
\mathcal{N}_{\mathrm{a}},
\]
and hence
\[
\omega
\left(
\mathcal{S}\cap\mathcal{S}'
\right)
\leq
\omega
\left(
\mathcal{N}_{\mathrm{a}}
\right)
\leq
\rho.
\]
This contradicts
\[
\omega
\left(
\mathcal{S}\cap\mathcal{S}'
\right)
\geq
\chi_{\boldsymbol{\omega}}(\eta)
>
\rho.
\]
Thus, two conflicting blocks cannot both be confirmed.

For necessity, suppose
$\chi_{\boldsymbol{\omega}}(\eta)\leq\rho$. Then there exist
$\mathcal{S},\mathcal{S}'\in\mathfrak{Q}_{\eta}$ such that
\[
\omega
\left(
\mathcal{S}\cap\mathcal{S}'
\right)
\leq
\rho.
\]
Let
$\mathcal{M}=\mathcal{S}\cap\mathcal{S}'$. The adversary may corrupt every
chain in $\mathcal{M}$.

Consider two conflicting branches containing blocks $B$ and $B'$. Chains in
$\mathcal{M}$ support both branches. Honest chains in
$\mathcal{S}\setminus\mathcal{M}$ support only $B$, while honest chains in
$\mathcal{S}'\setminus\mathcal{M}$ support only $B'$. The two honest sets are
disjoint, so no honest chain supports conflicting blocks. Before global
stabilization, the network schedule may delay information between the two
honest groups while each group extends its respective branch.

Since
$\omega(\mathcal{S})\geq\eta$ and
$\omega(\mathcal{S}')\geq\eta$, both branches can obtain confirmation
quorums. Therefore, safety against every corrupted set of weight at most
$\rho$ requires
$\chi_{\boldsymbol{\omega}}(\eta)>\rho$.

For liveness without Byzantine participation, if
$\omega(\mathcal{N}_{\mathrm{h}})\geq\eta$, then
$\mathcal{N}_{\mathrm{h}}\in\mathfrak{Q}_{\eta}$ and the honest chains can
form a confirmation quorum. Conversely, if
$\omega(\mathcal{N}_{\mathrm{h}})<\eta$, every exclusively honest set has
weight below $\eta$, so Byzantine participation is necessary. This proves
\eqref{eq:exact-quorum-liveness}.
\end{proof}

\begin{proof}[Proof of Corollary~\ref{cor:universal-threshold}]
For arbitrary
$\mathcal{S},\mathcal{S}'\in\mathfrak{Q}_{\eta}$,
\begin{align}
\omega
\left(
\mathcal{S}\cap\mathcal{S}'
\right)
&=
\omega(\mathcal{S})
+
\omega(\mathcal{S}')
-
\omega
\left(
\mathcal{S}\cup\mathcal{S}'
\right)
\nonumber\\
&\geq
2\eta-1,
\label{eq:proof-quorum-intersection-lower-bound}
\end{align}
because the total system weight is one. Hence,
\[
\chi_{\boldsymbol{\omega}}(\eta)
\geq
2\eta-1.
\]
If $\eta>(1+\rho)/2$, then
$2\eta-1>\rho$, and
Proposition~\ref{prop:tight-quorum} gives conflicting-block safety.

Moreover,
\[
\omega
\left(
\mathcal{N}_{\mathrm{h}}
\right)
=
1-
\omega
\left(
\mathcal{N}_{\mathrm{a}}
\right)
\geq
1-\rho.
\]
Thus, $\eta\leq1-\rho$ ensures that the honest chains can form a confirmation
quorum.

The interval
\[
\left(
\frac{1+\rho}{2},
1-\rho
\right]
\]
is nonempty if and only if
$(1+\rho)/2<1-\rho$, which is equivalent to
$\rho<1/3$.
\end{proof}

\begin{proof}[Proof of Corollary~\ref{cor:uniform-weight-threshold}]
With uniform weights, a confirmation quorum is any set of at least
$q_{\eta}=\lceil \eta N\rceil$ distinct chains. The minimum intersection
size of two subsets of $\{1,\ldots,N\}$ of size at least $q_{\eta}$ is
$\max\{0,2q_{\eta}-N\}$, and this minimum is achieved by two subsets of
size exactly $q_{\eta}$. Hence
\[
\chi_{\boldsymbol{\omega}}(\eta)
=
\frac{\max\{0,2q_{\eta}-N\}}{N}.
\]
The adversarial weight bound is $\rho=f/N$. Proposition~\ref{prop:tight-quorum}
therefore gives conflicting-block safety if and only if
\[
\frac{\max\{0,2q_{\eta}-N\}}{N}
>
\frac{f}{N},
\]
which is equivalent to $2q_{\eta}-N>f$. Honest-only confirmation is possible
if and only if the $N-f$ honest chains can form a confirmation quorum, namely
if and only if $q_{\eta}\leq N-f$.
\end{proof}

\section{Proof of the Cross-Layer DAG Stability Result}
\label{app:proof-dag-stability}

\begin{proof}[Proof of Theorem~\ref{thm:cross-layer-stability}]
For brevity, write
\[
\nu
=
\nu_{\mathrm{h}}\!\left(K,\mathbf{R}^{\mathrm{code}}\right).
\]
By independent thinning of the honest proposal processes,
$\widehat{H}_t$ is Poisson with mean $\nu$. By assumption, $A_t$ is Poisson
with mean $\lambda_{\mathrm{a}}$, and the two sequences are independent
across update intervals.

Fix $L_t=\ell\geq1$ and $\widehat{H}_t=h$. Let
$K_{\ell}=\min\{K,\ell\}$. For a fixed public tip, the probability of not
being selected by one honest block is
$1-K_{\ell}/\ell$. Conditional independence of the $h$ selections gives
\[
\Pr\!\left[
\text{the tip is not selected}
\mid
L_t=\ell,\widehat{H}_t=h
\right]
=
\left(
1-\frac{K_{\ell}}{\ell}
\right)^h.
\]
Introducing one indicator for each of the $\ell$ public tips and applying
linearity of expectation,
\[
\mathbb{E}
\left[
D_t
\mid
L_t=\ell,\widehat{H}_t=h
\right]
=
\ell
\left[
1-
\left(
1-\frac{K_{\ell}}{\ell}
\right)^h
\right].
\]

The probability-generating function of
$\widehat{H}_t\sim\operatorname{Poisson}(\nu)$ is
$\mathbb{E}[z^{\widehat{H}_t}]
=\exp(\nu(z-1))$. Therefore,
\begin{equation}
\mathbb{E}
\left[
D_t
\mid
L_t=\ell
\right]
=
\ell
\left[
1-
\exp\!\left(
-\frac{\nu K_{\ell}}{\ell}
\right)
\right].
\label{eq:proof-expected-tip-removals}
\end{equation}
Substituting
\eqref{eq:proof-expected-tip-removals} into
\eqref{eq:tip-process} yields
\[
\delta(\ell)
=
\nu+\lambda_{\mathrm{a}}
-
\ell
\left[
1-
\exp\!\left(
-\frac{\nu K_{\ell}}{\ell}
\right)
\right].
\]

For $\ell\geq K$, $K_{\ell}=K$, and
\[
\lim_{\ell\rightarrow\infty}
\ell
\left[
1-
\exp\!\left(
-\frac{\nu K}{\ell}
\right)
\right]
=
\nu K.
\]
Hence,
\begin{equation}
\lim_{\ell\rightarrow\infty}
\delta(\ell)
=
\lambda_{\mathrm{a}}
-
(K-1)\nu.
\label{eq:proof-asymptotic-drift}
\end{equation}

Suppose first that
$\lambda_{\mathrm{a}}<(K-1)\nu$. Choose $\varepsilon>0$ such that
\[
\lambda_{\mathrm{a}}-(K-1)\nu
<
-2\varepsilon.
\]
By~\eqref{eq:proof-asymptotic-drift}, there exists $\ell_0$ such that
\begin{equation}
\delta(\ell)
\leq
-\varepsilon,
\qquad
\ell\geq\ell_0.
\label{eq:proof-negative-linear-drift}
\end{equation}

Let $V_1(\ell)=\ell$. Equation
\eqref{eq:proof-negative-linear-drift} gives strictly negative drift outside
the finite set $\{0,\ldots,\ell_0-1\}$. The one-step increments have finite
first moment because
\[
D_t
\leq
K\widehat{H}_t
\]
and both $\widehat{H}_t$ and $A_t$ have finite moments. The
Foster--Lyapunov positive-recurrence criterion therefore implies that the
irreducible chain is positive recurrent. Irreducibility also gives uniqueness
of its stationary distribution.

To establish finiteness of the stationary first moment, define
$\Delta_t=L_{t+1}-L_t$. Since
$D_t\leq K\widehat{H}_t$,
\[
|\Delta_t|
\leq
(K+1)\widehat{H}_t+A_t.
\]
The Poisson moment bounds imply the existence of a finite constant $C_2$ such
that
\begin{equation}
\sup_{\ell\geq0}
\mathbb{E}
\left[
\Delta_t^2
\mid
L_t=\ell
\right]
\leq
C_2.
\label{eq:proof-increment-second-moment}
\end{equation}

For $V_2(\ell)=\ell^2$,
\begin{align}
\mathbb{E}
\left[
V_2(L_{t+1})-V_2(L_t)
\mid
L_t=\ell
\right]
&=
2\ell\delta(\ell)
+
\mathbb{E}
\left[
\Delta_t^2
\mid
L_t=\ell
\right]
\nonumber\\
&\leq
-2\varepsilon\ell+C_2
\end{align}
for $\ell\geq\ell_0$. Outside a sufficiently large finite set, the right-hand
side is at most $-\varepsilon\ell$. The moment form of the
Foster--Lyapunov criterion then gives
\[
\sum_{\ell=0}^{\infty}
\ell\pi(\ell)
<
\infty,
\]
where $\pi$ is the stationary distribution.

Now suppose that
$\lambda_{\mathrm{a}}>(K-1)\nu$. Since
$D_t\leq K\widehat{H}_t$,
\[
L_{t+1}-L_t
=
\widehat{H}_t+A_t-D_t
\geq
A_t-(K-1)\widehat{H}_t.
\]
Iteration gives
\[
L_t
\geq
L_0
+
\sum_{s=0}^{t-1}
\left[
A_s-(K-1)\widehat{H}_s
\right].
\]
The summands are independent and identically distributed with mean
$\lambda_{\mathrm{a}}-(K-1)\nu>0$. By the strong law of large numbers,
\[
\liminf_{t\rightarrow\infty}
\frac{L_t}{t}
\geq
\lambda_{\mathrm{a}}
-
(K-1)\nu
>
0
\]
almost surely. This proves
\eqref{eq:linear-tip-divergence} and rules out positive recurrence.

It remains to establish the explicit sufficient condition. From
\eqref{eq:completion-probability-lower-bound},
\[
\nu_{\mathrm{h}}\!\left(K,\mathbf{R}^{\mathrm{code}}\right)
\geq
\sum_{j\in\mathcal{N}_{\mathrm{h}}}
\lambda_j
\left[
\Psi_j\!\left(K,R_j^{\mathrm{code}}\right)
-
\epsilon_{\mathrm{ver},j}
\right]_{+}.
\]
Therefore, condition
\eqref{eq:explicit-cross-layer-stability} implies
\[
\lambda_{\mathrm{a}}
<
(K-1)
\nu_{\mathrm{h}}\!\left(K,\mathbf{R}^{\mathrm{code}}\right),
\]
and positive recurrence follows from the first part of the proof.

At the equality boundary, the asymptotic drift vanishes, and the preceding
Foster--Lyapunov and divergence arguments do not determine the recurrence
classification.
\end{proof}

\section{Proof of the Weighted Confirmation-Time Law}
\label{app:proof-confirmation-time}

\begin{proof}[Proof of Proposition~\ref{prop:weighted-confirmation-time}]
For integer $r\geq0$, define
\[
\mathcal{S}_r(B)
=
\left\{
j\in\mathcal{N}_{\mathrm{h}}:
\tau_j(B)\leq r
\right\}.
\]
By definition of $T_{\mathrm{dag}}(B)$,
\[
\left\{
T_{\mathrm{dag}}(B)>r
\right\}
=
\left\{
\omega
\left(
\mathcal{S}_r(B)
\right)
<
\eta
\right\}.
\]

Fix
$\mathcal{S}\subseteq\mathcal{N}_{\mathrm{h}}$. Independence of the support
times gives
\begin{align}
\Pr\!\left[
\mathcal{S}_r(B)=\mathcal{S}
\right]
&=
\prod_{j\in\mathcal{S}}
\Pr\!\left[
\tau_j(B)\leq r
\right]
\prod_{j\in\mathcal{N}_{\mathrm{h}}\setminus\mathcal{S}}
\Pr\!\left[
\tau_j(B)>r
\right]
\nonumber\\
&=
\prod_{j\in\mathcal{S}}
\left[
1-
\left(
1-p_j^{\mathrm{sup}}(B)
\right)^r
\right]
\prod_{j\in\mathcal{N}_{\mathrm{h}}\setminus\mathcal{S}}
\left(
1-p_j^{\mathrm{sup}}(B)
\right)^r.
\label{eq:proof-support-set-law}
\end{align}
The events
$\{\mathcal{S}_r(B)=\mathcal{S}\}$ partition the sample space. Summing
\eqref{eq:proof-support-set-law} over all sets satisfying
$\omega(\mathcal{S})<\eta$ proves
\eqref{eq:exact-confirmation-tail}.

Since $T_{\mathrm{dag}}(B)$ is a nonnegative integer-valued random variable,
the tail-sum identity gives
\[
\mathbb{E}
\left[
T_{\mathrm{dag}}(B)
\right]
=
\sum_{r=0}^{\infty}
\Pr\!\left[
T_{\mathrm{dag}}(B)>r
\right].
\]

Let $\mathcal{H}\subseteq\mathcal{N}_{\mathrm{h}}$ be an honest confirmation
quorum such that
$\omega(\mathcal{H})\geq\eta$ and
$p_j^{\mathrm{sup}}(B)>0$ for all $j\in\mathcal{H}$. Once every chain in
$\mathcal{H}$ has supported $B$, the block is confirmed. Hence,
\[
T_{\mathrm{dag}}(B)
\leq
\max_{j\in\mathcal{H}}
\tau_j(B)
\leq
\sum_{j\in\mathcal{H}}
\tau_j(B).
\]
Since
$\mathbb{E}[\tau_j(B)]=1/p_j^{\mathrm{sup}}(B)$,
\[
\mathbb{E}
\left[
T_{\mathrm{dag}}(B)
\right]
\leq
\sum_{j\in\mathcal{H}}
\frac{1}{p_j^{\mathrm{sup}}(B)}
<
\infty.
\]
\end{proof}

\section{Proof of the Cross-Layer Settlement Guarantee}
\label{app:proof-cross-layer-settlement}

\begin{proof}[Proof of Theorem~\ref{thm:cross-layer-settlement}]
For validation instance $h$, let
$\mathcal{B}_{\mathrm{bad}}^{(h)}=\mathcal{E}_{\mathrm{val}}^{(h)}$.
By Proposition~\ref{prop:polar-validation-soundness},
\[
\Pr\!\left[
\mathcal{B}_{\mathrm{bad}}^{(h)}
\right]
\leq
\epsilon_{\mathrm{ver}}^{(h)}.
\]
Define
\[
\mathcal{B}_{\mathrm{bad}}^{(H)}
=
\bigcup_{h=1}^{H}
\mathcal{B}_{\mathrm{bad}}^{(h)}.
\]
The union bound gives
\begin{equation}
\Pr\!\left[
\mathcal{B}_{\mathrm{bad}}^{(H)}
\right]
\leq
\sum_{h=1}^{H}
\epsilon_{\mathrm{ver}}^{(h)}.
\label{eq:proof-global-validation-error}
\end{equation}

We show that
\begin{equation}
\mathcal{E}_{\mathrm{set}}^{(H)}
\subseteq
\mathcal{B}_{\mathrm{bad}}^{(H)}.
\label{eq:proof-settlement-event-inclusion}
\end{equation}
Condition on
$(\mathcal{B}_{\mathrm{bad}}^{(H)})^{\mathrm{c}}$. Every validation result
accepted by an honest committee is then numerically correct.

First, every confirmation quorum contains an honest chain. Let
$\mathcal{S}\in\mathfrak{Q}_{\eta}$ be a confirmation quorum. Since the pair
$(\mathcal{S},\mathcal{S})$ is included in the minimization defining
$\chi_{\boldsymbol{\omega}}(\eta)$,
\[
\omega(\mathcal{S})
\geq
\chi_{\boldsymbol{\omega}}(\eta)
>
\rho.
\]
The Byzantine chains have total weight at most $\rho$, so
$\mathcal{S}\nsubseteq\mathcal{N}_{\mathrm{a}}$.

Consequently, every confirmed block is contained in the accepted past cone
of at least one honest supporting chain. Before contributing support, that
chain independently verifies the block's deterministic admissibility against
its referenced checkpoint, including the source reservation, issuer sequence
number, replay identifiers, nonces, sufficient-funds condition, validation
certificate, and conflict-freedom of the relevant past cone.

Order the blocks in $\mathcal{F}_H$ according to the deterministic checkpoint
order:
\[
B^{(1)}
\prec_{\mathrm{cp}}
\cdots
\prec_{\mathrm{cp}}
B^{(m)}.
\]
Let $\mathbf{b}^{(r)}$ be the state after applying the first $r$ blocks.

We prove by induction that
$\mathbf{b}^{(r)}\succeq\mathbf{0}$ for
$r=0,\ldots,m$. The claim holds at $r=0$ by initialization. Consider
$B^{(r)}$, issued by chain $j$, and let
$\mathbf{b}^{\mathrm{cp}}$ be its referenced checkpoint state. Honest support
implies
\begin{equation}
\mathbf{b}^{\mathrm{cp}}
-
\mathbf{d}^{(r)}
\succeq
\mathbf{0}.
\label{eq:proof-checkpoint-admissibility}
\end{equation}

Only chain $j$ may debit coordinates in $\mathcal{U}_j$. By the
one-outstanding-sequence restriction, no other block from chain $j$ validated
against the same checkpoint can precede $B^{(r)}$ in the application order;
a later sequence from the same issuer can be formed only after the preceding
outgoing sequence has been resolved in the checkpoint state. Blocks issued
by other chains have zero debit on $\mathcal{U}_j$ and may only add credits
to those coordinates. Updates from earlier checkpoint intervals are already
included in $\mathbf{b}^{\mathrm{cp}}$. Therefore, on every coordinate in
$\operatorname{supp}(\mathbf{d}^{(r)})$,
\[
\mathbf{b}^{(r-1)}
\succeq
\mathbf{b}^{\mathrm{cp}}.
\]
Together with
\eqref{eq:proof-checkpoint-admissibility}, this gives
\[
\mathbf{b}^{(r-1)}
-
\mathbf{d}^{(r)}
\succeq
\mathbf{0}.
\]
Since
$\mathbf{c}^{(r)}\succeq\mathbf{0}$,
\[
\mathbf{b}^{(r)}
=
\mathbf{b}^{(r-1)}
-
\mathbf{d}^{(r)}
+
\mathbf{c}^{(r)}
\succeq
\mathbf{0}.
\]
Thus,
$\mathcal{E}_{\mathrm{neg}}^{(H)}$ does not occur.

For every transfer block,
\[
\mathbf{1}^{\mathsf T}\mathbf{d}^{(r)}
=
\mathbf{1}^{\mathsf T}\mathbf{c}^{(r)}.
\]
Hence,
\[
\mathbf{1}^{\mathsf T}\mathbf{b}^{(r)}
=
\mathbf{1}^{\mathsf T}\mathbf{b}^{(r-1)}.
\]
Induction yields
\[
\mathbf{1}^{\mathsf T}\mathbf{b}^{(r)}
=
\mathbf{1}^{\mathsf T}\mathbf{b}^{(0)}
\]
for every $r$. Therefore,
$\mathcal{E}_{\mathrm{con}}^{(H)}$ does not occur. Source reservations and
destination-chain realizations do not alter this equality because neither
operation applies an additional update to the abstract settlement state.

By~\eqref{eq:exact-quorum-safety} and
Proposition~\ref{prop:tight-quorum}, conflicting blocks cannot both obtain
confirmation quorums. Therefore,
$\mathcal{E}_{\mathrm{conf}}^{(H)}$ does not occur.

For each transaction identifier $\iota$, the destination adapter executes an
atomic check-record-apply transition. If $\iota$ is absent from the replay
state, the adapter records it and applies the destination credit in the same
atomic transition. If $\iota$ is already present, the transfer is ignored.
Consequently,
$N_H(\iota)\leq1$ for every $\iota$, and
$\mathcal{E}_{\mathrm{dup}}^{(H)}$ does not occur.

It remains to verify coded-state consistency. Consider an honest chain
$j\in\mathcal{N}_{\mathrm{h}}$ and an honest worker
$i\in\mathcal{W}_j(t)$. During an ordinary checkpoint transition, the worker
accepts only the committee-certified increment
$\Delta\widetilde{\mathbf{b}}_{j,i}^{\mathrm{conf}}(t)$. Assuming the coded
invariant at checkpoint $t$,
\begin{align}
\widetilde{\mathbf{b}}_{j,i}(t+1)
&=
\widetilde{\mathbf{b}}_{j,i}(t)
+
\Delta\widetilde{\mathbf{b}}_{j,i}^{\mathrm{conf}}(t)
\nonumber\\
&=
\sum_{\ell=1}^{k_j(t)}
\mathbf{G}_j(t)[i,\ell]
\left(
\mathbf{b}_{j,\ell}(t)
+
\Delta\mathbf{b}_{j,\ell}^{\mathrm{conf}}(t)
\right)
\nonumber\\
&=
\sum_{\ell=1}^{k_j(t)}
\mathbf{G}_j(t)[i,\ell]
\mathbf{b}_{j,\ell}(t+1).
\end{align}
At a reconfiguration checkpoint, the honest committee re-encodes the
confirmed raw state under the new generator matrix before activating the new
configuration. Induction over checkpoints proves the invariant for all
honest workers of honest chains. Hence,
$\mathcal{E}_{\mathrm{code}}^{(H)}$ does not occur.

We have shown that
$(\mathcal{B}_{\mathrm{bad}}^{(H)})^{\mathrm{c}}$ implies
$(\mathcal{E}_{\mathrm{set}}^{(H)})^{\mathrm{c}}$, proving
\eqref{eq:proof-settlement-event-inclusion}. Combining
\eqref{eq:proof-settlement-event-inclusion} with
\eqref{eq:proof-global-validation-error} gives
\[
\Pr\!\left[
\mathcal{E}_{\mathrm{set}}^{(H)}
\right]
\leq
\sum_{h=1}^{H}
\epsilon_{\mathrm{ver}}^{(h)}.
\]

For liveness, let
$\mathcal{H}\subseteq\mathcal{N}_{\mathrm{h}}$ be an honest confirmation
quorum satisfying
$\omega(\mathcal{H})\geq\eta$ and
$p_j^{\mathrm{sup}}(B)>0$ for every $j\in\mathcal{H}$. Under the geometric
support model,
\[
\Pr\!\left[
\tau_j(B)<\infty
\right]
=
1
\]
for every $j\in\mathcal{H}$. Because $\mathcal{H}$ is finite,
\[
\max_{j\in\mathcal{H}}
\tau_j(B)
<
\infty
\]
almost surely. Once every chain in $\mathcal{H}$ has supported $B$, the
accumulated honest weight is at least $\eta$. Thus,
\[
T_{\mathrm{dag}}(B)
\leq
\max_{j\in\mathcal{H}}
\tau_j(B)
<
\infty
\]
almost surely.

Furthermore,
\[
\mathbb{E}
\left[
T_{\mathrm{dag}}(B)
\right]
\leq
\sum_{j\in\mathcal{H}}
\frac{1}{p_j^{\mathrm{sup}}(B)}
<
\infty.
\]
The DAG-stability condition places this support process in the stable
operating regime in which the assumed positive support probabilities remain
compatible with bounded public-tip backlog.

Finally,
\[
T_{\mathrm{fin}}(B)
=
T_{\mathrm{val}}(B)
+
T_{\mathrm{dag}}(B)
+
T_{\mathrm{apply}}(B).
\]
Therefore, finite means for
$T_{\mathrm{val}}(B)$ and
$T_{\mathrm{apply}}(B)$, together with the finite confirmation-time bound
above, imply
\[
\mathbb{E}
\left[
T_{\mathrm{fin}}(B)
\right]
<
\infty.
\]
\end{proof}

\bibliographystyle{elsarticle-num}
\bibliography{refs}
\end{document}